\documentclass[acmtog]{acmart} %
\acmDOI{10.1145/3763305}
\usepackage{amsmath,amsthm,amssymb}
\usepackage{amsfonts}
\usepackage{xcolor}
\usepackage{algorithm}
\usepackage{soul}
\usepackage{xspace}
\usepackage{bm}
\usepackage{fancybox}

\usepackage{algorithm}
\usepackage{algorithmic}
\newcommand{\FullComment}[1]{\STATE{\color{primalComment}$\triangleright$~#1}}

\DeclareUnicodeCharacter{0302}{\^{}}

\newcommand{\primalHue}{30}
\newcommand{\differentialHue}{230}
\newcommand{\backgroundSaturation}{0.1}
\newcommand{\foregroundSaturation}{0.2}
\newcommand{\commentSaturation}{0.8}
\newcommand{\brightness}{1}

\definecolor{primalBackground}{Hsb}{\primalHue, \backgroundSaturation, \brightness}
\colorlet{primalBackground}[rgb]{primalBackground}
\definecolor{primalForeground}{Hsb}{\primalHue, \foregroundSaturation, \brightness}
\colorlet{primalForeground}[rgb]{primalForeground}
\definecolor{primalComment}{Hsb}{\primalHue, \commentSaturation, \brightness}
\colorlet{primalComment}[rgb]{primalComment}
\definecolor{differentialBackground}{Hsb}{\differentialHue, \backgroundSaturation, \brightness}
\colorlet{differentialBackground}[rgb]{differentialBackground}
\definecolor{differentialForeground}{Hsb}{\differentialHue, \foregroundSaturation, \brightness}
\colorlet{differentialForeground}[rgb]{differentialForeground}
\definecolor{differentialComment}{Hsb}{\differentialHue, \commentSaturation, \brightness}
\colorlet{differentialComment}[rgb]{differentialComment}

\makeatletter
\newlength{\titledeqpad} %
\newcommand{\titledeq}[4]{%
  \par\medskip
  \begingroup
    \setlength{\fboxrule}{1.5pt}%
    \setlength{\fboxsep}{0pt}%
    \noindent
    \fcolorbox{#2}{#3}{%
      \parbox{\linewidth}{%
        \begingroup
          \setlength{\fboxsep}{0pt}%
          \colorbox{#2}{%
            \parbox{\linewidth}{%
              \vspace{2pt}%
              \hspace{2pt}{\normalsize\textsc{#1}}\hspace{2pt}%
              \vspace{3pt}%
            }%
          }%
        \endgroup
        \par\vspace{2pt}%
        \noindent\hspace*{\titledeqpad}%
        \begin{minipage}{\dimexpr\linewidth-2\titledeqpad\relax}
          \normalsize #4%
        \end{minipage}%
        \hspace*{\titledeqpad}\par\vspace{3pt}%
      }%
    }%
  \endgroup
  \par\medskip
}
\makeatother

\cornersize*{0.22pt}

\newsavebox{\mcbContent}
\newsavebox{\mcbTitle}
\newsavebox{\mcbInner}
\newlength{\mcbW}
\newlength{\mcbPad}
\newcommand{\mathcalloutbox}[4]{%
  \mathrel{%
    \begingroup
      \sbox{\mcbContent}{$\displaystyle #4$}%
      \sbox{\mcbTitle}{\scriptsize\bfseries #1}%
      \setlength{\mcbW}{\wd\mcbContent}%
      \ifdim\wd\mcbTitle>\mcbW \setlength{\mcbW}{\wd\mcbTitle}\fi
      \setlength{\fboxsep}{\mcbPad}%
      \sbox{\mcbInner}{%
        \vbox{\offinterlineskip
          \hbox{\colorbox{white}{\makebox[\mcbW]{\color{black}\scriptsize\bfseries #1}}}%
          \vskip-0.2pt %
          \hbox{\colorbox{#2}{\makebox[\mcbW]{\color{black}$\displaystyle #4$}}}%
        }%
      }%
      \leavevmode
      \hbox{%
        \usebox{\mcbInner}%
        \kern-\wd\mcbInner
        {\color{#3}\setlength{\fboxsep}{0pt}\ovalbox{\phantom{\usebox{\mcbInner}}}}%
      }%
    \endgroup
  }%
}

\acmJournal{TOG}

\begin{document}
\title{Differentiable Light Transport with Gaussian Surfels via Adapted Radiosity for Efficient Relighting and Geometry Reconstruction}

\author{Kaiwen Jiang}
\email{k1jiang@ucsd.edu}
\affiliation{%
 \institution{University of California, San Diego}
 \country{United States of America}
}

\author{Jia-Mu Sun}
\email{jis081@ucsd.edu}
\affiliation{%
 \institution{University of California, San Diego}
 \country{United States of America}
 }

\author{Zilu Li}
\email{zil124@ucsd.edu}
\affiliation{%
 \institution{University of California, San Diego}
 \country{United States of America}
 }

\author{Dan Wang}
\email{danwang@ucsd.edu}
\affiliation{%
 \institution{University of California, San Diego}
 \country{United States of America}
}

\author{Tzu-Mao Li}
\email{tzli@ucsd.edu}
\affiliation{%
 \institution{University of California, San Diego}
 \country{United States of America}
}
 
\author{Ravi Ramamoorthi}
\email{ravir@cs.ucsd.edu}
\affiliation{
 \institution{University of California, San Diego}
 \country{United States of America}
}

\authorsaddresses{}

\begin{teaserfigure}
\begin{flushleft}
\vspace{-0.3cm}
{\large \textcolor{magenta}{\texttt{\href{https://raymondjiangkw.github.io/radiositygs.github.io/}{https://raymondjiangkw.github.io/radiositygs.github.io/}}}}\\
\vspace{0.1cm}
\begin{center}
    \centering
    \includegraphics[width=\linewidth]{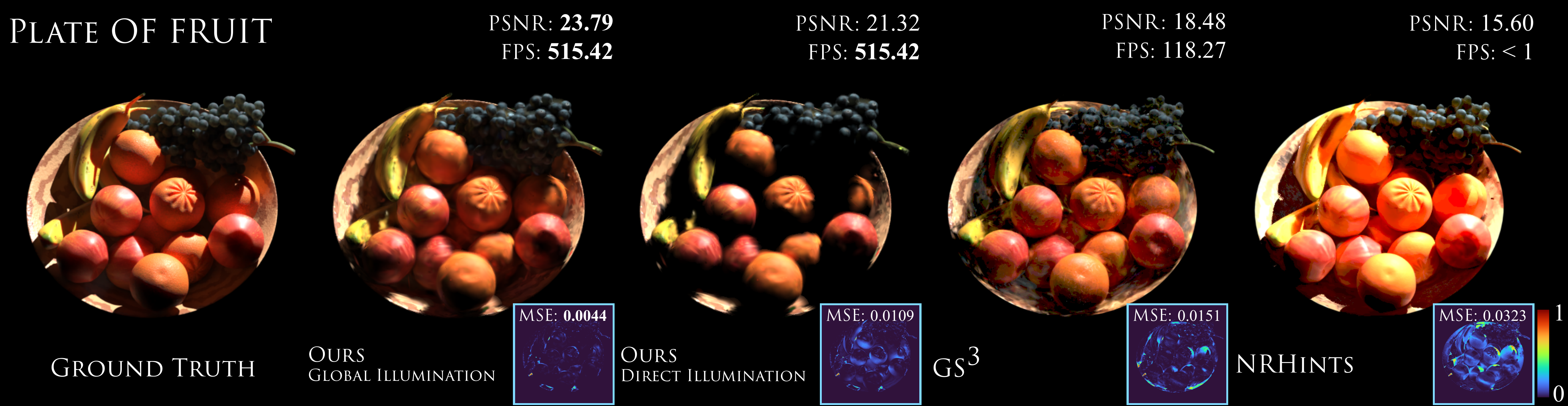}
    \caption{
    We present a differentiable light transport framework with Gaussian surfels as primitives for efficient relighting and geometry reconstruction. Given $100$ images with varying known point light sources, varying viewpoints and random initialization, our method outperforms the baselines GS$^3$ \cite{gs3} and NRHints \cite{nrhints} at the test view with unseen viewpoint and lighting conditions, achieving high-quality relighting and extremely fast novel view synthesis as evidenced by the PSNR and FPS metrics shown in the top-right corner. In our method, when only direct illumination is considered, the inter-reflection among fruits and color bleeding effects are compromised, leading to lower view synthesis quality. 
    We calculate the error maps between the rendered results and the ground truth using the MSE metric, displayed as insets in the bottom-right corner. The color bar is put at the right. Ours with global illumination achieves the lowest MSE, while others contain prominent errors for shadows or indirect illumination effects.
    We further demonstrate the applications of our method in Fig.~\ref{fig:teaser-2}.
    The metrics are measured on an NVIDIA 6000 Ada GPU.
    Scene adapted from \textsc{fresh fruits in a bowl} by Sougatadhar16\textcopyright TurboSquid.
    }
    \label{fig:teaser}
\end{center}
\end{flushleft}
\end{teaserfigure}

\begin{abstract}
Radiance fields have gained tremendous success with applications ranging from novel view synthesis to geometry reconstruction, especially with the advent of Gaussian splatting. However, they sacrifice modeling of material reflective properties and lighting conditions, leading to significant geometric ambiguities and the inability to easily perform relighting. 
One way to address these limitations is to incorporate physically-based rendering, but it has been prohibitively expensive to include full global illumination within the inner loop of the optimization. Therefore, previous works adopt simplifications that make the whole optimization with global illumination effects efficient but less accurate. 
In this work, we adopt Gaussian surfels as the primitives and build an efficient framework for differentiable light transport, inspired from the classic radiosity theory. The whole framework operates in the coefficient space of spherical harmonics, enabling both diffuse and specular materials. We extend the classic radiosity into non-binary visibility and semi-opaque primitives, propose novel solvers to efficiently solve the light transport, and derive the backward pass for gradient optimizations, which is more efficient than auto-differentiation. 
During inference, we achieve \emph{view-independent} rendering where light transport need not be recomputed under viewpoint changes, enabling hundreds of FPS for global illumination effects, including view-dependent reflections using a spherical harmonics representation.
Through extensive qualitative and quantitative experiments, we demonstrate superior geometry reconstruction, view synthesis and relighting than previous inverse rendering baselines, or data-driven baselines given relatively sparse datasets with known or unknown lighting conditions.
\end{abstract}


\maketitle

\section{Introduction}

The success of radiance fields has enabled a broad range of applications ranging from novel view synthesis \cite{nerf2020} to geometry reconstruction \cite{neus}. Compared to the classic rendering algorithm which parametrizes lighting, geometry, materials, \emph{etc.}, the radiance fields directly parametrize the outgoing radiance, which greatly simplifies forward and inverse rendering and facilitates optimization. Typically, existing popular methods represent radiance fields in two distinct ways: 1) neural-based continuous representation (e.g., \cite{nerf2020, instant-ngp, nerfstudio}); 2) kernel-based discrete representation (e.g., \cite{3DGS, 2DGS, deformableKernel}).

However, despite the success of radiance fields in view synthesis, they sacrifice the editability of the underlying materials, increase geometric ambiguity, and are unable to perform relighting.

Observing such limitations, the well-known solution is to incorporate physically-based rendering with global illumination. However, it is notoriously costly especially when used in the inner loop of an inverse rendering algorithm. 
Within the neural-based continuous representation, representative works (e.g., \cite{nerv, nrtf, flashcache, tensoir}) adopt simplifications (such as considering only single bounce, ignoring visibility, initializing from radiance fields, \emph{etc.}) to improve the efficiency. 
Within the kernel-based discrete representation, existing methods (e.g., \cite{Gaussianshader, relightable3DGS, GSIR, GIGS}) also adopt similar simplifications, including learning the indirect illumination as separated outgoing radiances, freezing the geometry or binarizing the visibility, to deal with datasets under static illumination conditions. Even for direct illumination, existing works usually only solve for far-field light sources, i.e., environment maps.
A different category of data-driven methods (e.g., \cite{nrhints, gs3, olatGaussian}) proposes to use a neural network to compress all the lighting details using datasets with varying illumination conditions, but these methods do not have the full understanding of the light transport and therefore require very densely captured OLAT (one light at a time) datasets, with typically more than $500$ images, for training.

\begin{figure}[t]
    \centering
    \includegraphics[width=\linewidth]{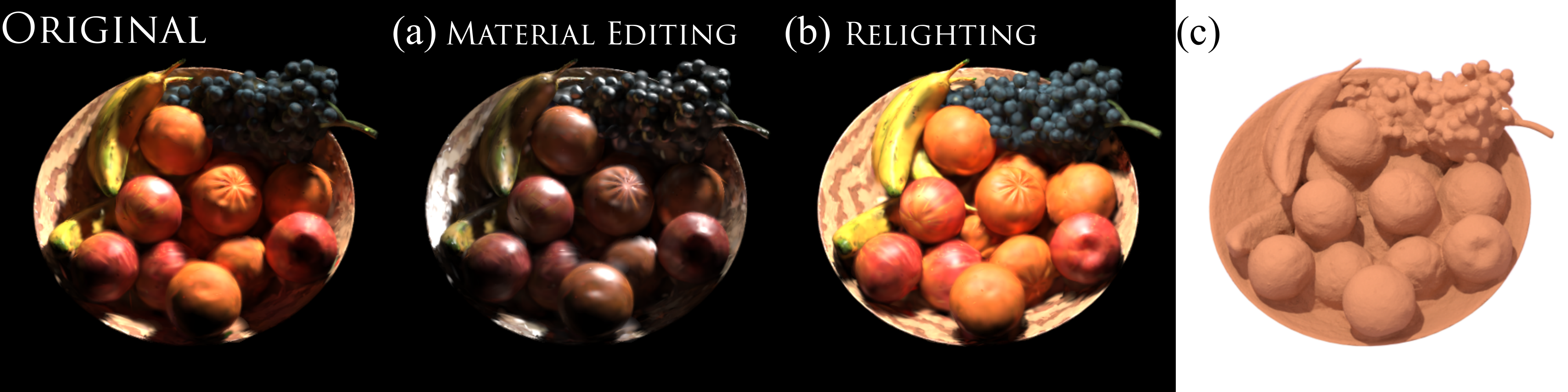}
    \caption{Demonstration of applications enabled by our method. Given the original result on the left, our method enables \textbf{(a)} material editing where we make the surface more shiny and set the specular albedo to be white, \textbf{(b)} relighting, and \textbf{(c)} geometry reconstruction.}
    \label{fig:teaser-2}
\end{figure}

We address the above-mentioned limitations, presenting the first efficient framework using Gaussian surfels \cite{2DGS, gfsgs} to differentiably calculate the physically based light transport for {relatively diffuse (but not pure Lambertian)} materials and non-emissive volumes. Our method enables applications including material editing, relighting and accurate geometry reconstruction as shown in Fig.~\ref{fig:teaser-2}. 
We solve both direct illumination and indirect illumination for near-field and far-field light sources, which has not been demonstrated in previous works.
As shown in Fig.~\ref{fig:teaser}, given relatively sparse inputs, physically based differentiable rendering significantly boosts the relighting quality compared to the data-driven approaches even with the direct illumination only. However, only using direct illumination omits relatively subtle but still vital indirect illumination effects. %

We choose Gaussian surfels for their efficiency and accurate definition of geometry \cite{gfsgs}. Within the context of Gaussian splatting, \citet{zhou2024unified} enables the forward light transport using ray tracing for 3D Gaussians primitives and several works (e.g., \cite{dont-splat-your-Gaussians, raysplats, 3dGaussianraytracing, 3dgut}) enable differentiable ray tracing under the radiance fields assumption. However, there are no existing works truly solving the differentiable global illumination, especially for Gaussian surfels. 
Even though ray tracing is pervasive, it is costly to iteratively traverse the scene and difficult to tradeoff between time consumption and variance, which in the end usually requires a denoiser. These factors prohibit robust optimization.

In contrast, the choice of using Gaussian surfels prompts us to revisit the classical theory of radiosity~\cite{radiosity} and introduce a radiosity-inspired method. Our method turns out to be very efficient for solving the light transport between Gaussian surfels, while maintaining low variance of gradients for robust optimization. Specifically, since Gaussian surfels are already discrete finite elements, we use the finite element radiosity framework that has been {well-developed} in computer graphics in the 1980s and 90s, but adapt it in significant ways for volumetric light transport through the Gaussian surfel primitives. We also consider differentiable rendering, as needed for the inverse problem, which has not previously been considered in a radiosity method.

We note that light transport in Gaussian surfels is ideally suited to a radiosity or finite element method, since no additional discretization step or refinement is needed as in conventional mesh-based radiosity. Moreover, visibility is continuous and non-binary, which eliminates the need for complicated meshing techniques~\cite{fredo_thesis,discontinuity-meshing,visibility_skeleton}. We are interested in relatively diffuse (but not pure Lambertian) light transport, which can be handled by low-order (we use $9$ in practice) spherical harmonics as in traditional radiosity methods \cite{radiosity_sh}. {Therefore, view-dependent effects are supported.}

In our particular application, finite element or radiosity approaches have the advantage of solving for all Gaussian surfels \emph{simultaneously}, for which we propose efficient parallel-friendly novel solvers. In contrast, in classical path tracing, we solve for each pixel independently and within each pixel, every path is independent from each other, leading to sub-optimal convergence speed.
Finite element or radiosity approaches also support \emph{view-independent} rendering with an iterative solution free of characteristic Monte-Carlo noise in ray tracing, which {saves the recomputation of light transport when viewpoint changes and therefore} enables %
hundreds of FPS during inference for global illumination effects as shown in Fig.~\ref{fig:teaser}. Finally, we also present necessary approximations that facilitate the optimization and greatly speed up the whole procedure.

Our main contributions are summarized as follows:
\begin{itemize}
    \item We present a forward rendering framework, inspired from the radiosity theory, with both direct illumination and indirect illumination for Gaussian surfels (Sec.~\ref{sec:forward-rendering}), that supports \emph{view-independent} rendering {and \emph{view-dependent} effects}.
    \item We deduce efficient gradient calculation for differentiable rendering in our framework (Sec.~\ref{method-backward}), achieving $\sim10\times$ speedup and significantly less memory footprint than automatic differentiation (Sec.~\ref{sec-exp-ablation}).
    \item We present novel solvers for efficiently solving for global illumination (Sec.~\ref{method-solver}) and useful approximations to facilitate the optimization (Sec.~\ref{method-approx}).
    \item We achieve high-quality relighting results with global illumination effects (Figs~\ref{fig:teaser},\ref{fig:view-synthesis}) and accurate geometry reconstruction (Fig.~\ref{fig:stanford-orb}) on common benchmarks.
\end{itemize}
\section{Related Work}
Our paper brings together and significantly extends many lines of work, including relighting and geometry reconstruction with radiance field methods, differentiable rendering with global illumination, and classical radiosity approaches.  

\paragraph{Relighting and geometry reconstruction in radiance fields.} Radiance fields combined with volume rendering have obtained significant success in recent years, which could be traced back to early explorations of view synthesis \cite{related_classic_work_3, related_classic_work_1, related_classic_work_7, related_classic_work_2, related_classic_work_6}. 
Many works explore the best primitives to represent the radiance fields, including continuous representations such as pure MLPs \cite{nerf2020, barron2021mip, barron2022mip}, hash tables \cite{instant-ngp, zip-nerf}, voxels \cite{plenoxels, liu2020neural}, octrees \cite{plenoctrees}, triplanes \cite{eg3d}, \emph{etc.} and discrete representations such as 2D/3D Gaussian kernels \cite{3DGS, 2DGS, Gaussian_surfel, volume-splatting},  Epanechnikov kernels \cite{dont-splat-your-Gaussians}, deformable radial kernels \cite{deformableKernel}, \emph{etc.}
With their powerful capability of view synthesis, the next question is regarding relighting and geometry reconstruction.

Relighting is typically solved by integrating physically-based rendering, which can be challenging due to its computational complexity. The integration is usually achieved by using ray tracing with different simplifications (e.g., \cite{nerv, nrtf, Gaussianshader, relightable3DGS, GSIR, GIGS, relightable_avatar}), such as limiting the indirect illumination, and ignoring, freezing or binarizing the visibility. Jointly training with continuous radiance fields (e.g., \cite{tensoir, flashcache}) strikes a better balance between accuracy and efficiency, but for discrete primitives, efficiently solving the physically-based rendering remains an open problem. 
Another category of data-driven methods (e.g., \cite{nrhints, gs3, olatGaussian}) proposes to use neural networks to learn all the lighting details using various hints, and then achieve high-quality relighting during test time. However, they require very densely captured OLAT (One-Light-At-a-Time) datasets, which may not always be available. 
Compared to them, our method is built upon discrete primitives, i.e., Gaussian surfels, which are well-suited for a different category of light transport calculation, i.e., finite element methods or radiosity. We do not compromise the indirect illumination or visibility and achieve efficient full differentiation.

Geometry reconstruction from images is also a topic gaining broad interest, which could be dated back to multi-view stereo methods (e.g., \cite{multi-view-stereo, multi-view-stereo-2, 4359315, 5226635, 5989831, 1206509, Bleyer2011PatchMatchS}). Built upon continuous representations of radiance fields, some methods propose to define a geometry using distance fields (e.g., \cite{neus, neus2, long2022sparseneus, volsdf, neuralangelo, object_as_volumes, wang2025unisdf, neudf, neuraludf, zimo2025hotspot}) and then convert them into density fields for differentiable volume rendering. Built upon the discrete representations of radiance fields, many methods propose to approximate a geometry from the density fields (e.g., \cite{guedon2023sugar, 2DGS, Gaussian_surfel, radegs, GOF, pgsr}). Some methods~\cite{gfsgs, dipole} define a stochastic geometry field~\cite{object_as_volumes} using discrete primitives and then convert it into density fields for differentiable volume rendering. \citet{many-world} proposes to differentiate a volumetric perturbation of a surface. However, geometry reconstruction based on the radiance fields or the \emph{emissive surfaces/volumes assumption} dramatically increases the ambiguity of the underlying shape, which is most evident on texture-less or heavily shadowed areas. Compared to these approaches, our method is based on accurate definition of the geometry using Gaussian surfels~\cite{gfsgs}, and enables the light transport calculation in the optimization to explain the texture-less or heavily shadowed areas. This requires awareness of light sources, but allows for more accurate geometry reconstruction.
\paragraph{Differentiable global illumination.} Early approaches of inverse (global) illumination (e.g., \cite{single-image-inverse-rendering, inverse_global, shape_interreflections, rama-paper}) often come with various constraints. Recently, differentiating physically-based rendering in ray or path tracing, combined with gradient-based optimization, {started} to show promise and generality~\cite{azinovic2019inverse,zhang2023neilf++,wu2023nefii,sun2023neural,wu2023factorized}. 
In the case of surface rendering, a main challenge comes from the discontinuous boundaries which is addressed by previous work~\cite{differentiable_monte_carlo, bangaru2020warpedsampling, Xu:2023:PSDR-WAS, reparameterize_discontinuous_integrands}. 
\citet{adjoint_differentiable} and \citet{PathReplay} formulate the derivatives of the rendering equation as an adjoint integral which could be efficiently calculated. 
Another line of work~\cite{belhe2024importance, mc_differentiable, fischer2023plateau} proposes tailored importance sampling strategies for gradient calculation. 
Some earlier pioneering work focused on volume rendering~\cite{volume_rendering_optim_first, shuang_inverse_volume, gkioulekas2013inverse, ioannis_inverse_rendering}, while later work~\cite{unbiased_inverse_volume_rendering, Zhang:2020:PSDR,differential_theory_radiative_transfer, path_space_differentiable} achieves full optimization of geometry and materials and address gradient bias and variance issues.

However, there has been no existing work that addresses the differentiable light transport for discrete primitives. Our differentiation of global illumination is studied using the finite element method, different from the established works using ray or path tracing. We deal with continuous non-binary visibility and our method therefore is free of the discontinuity issues. We also develop a novel derivation to calculate the gradients of geometry and materials in a more efficient way than automatic differentiation. 
\paragraph{Radiosity.} 
The computer graphics community has studied realistic image synthesis for decades, with the radiosity method \cite{first_radiosity} %
introduced as an alternative to ray tracing \cite{first_ray_tracing} and raster graphics. Both radiosity and ray tracing approaches are later reformulated to solve the rendering equation \cite{rendering_equation}.
Compared to path tracing, which uses Monte-Carlo estimation to iteratively solve the recursive integral from the camera plane, radiosity divides the {structured} geometry into many \emph{elements}, and explicitly models the light transport between elements. The light transport is then formulated as a linear system, whose solution solves the rendering equation. Notably, the radiosity approach achieves \emph{view-independent} rendering in contrast to \emph{view-dependent} rendering of ray tracing. 
The initial radiosity approach is limited to the Lambertian surfaces \cite{hemi-cube}, while it is later extended to non-diffuse surfaces \cite{radiosity-non-diffuse} as well using two-pass methods \cite{radiosity-two-pass, 10.1145/378456.378492}, spherical harmonics \cite{radiosity_sh}, \emph{etc.} 
Various solvers (such as Jacobi iteration, \cite{radiosity_progressive_refinement}) are introduced to efficiently solve the linear system from the shooting or gathering perspective, while hierarchical methods \cite{radiosity_hierarchical, radiosity_hierarchical_glossy} are introduced to systematically reduce the complexity.
However, the radiosity approach is gradually replaced by the ray tracing approach and receives less and less attention over the years due to its discontinuity meshing issues for visibility \cite{fredo_thesis,discontinuity-meshing}, incapability of handling very complex materials, memory and time complexity of maintaining and solving the linear system. Hadadan et al.~\shortcite{neural_radiosity,hadadan2022differentiable,hadadan2023inverse} recently combined radiosity with neural networks to effectively solve the rendering equation and its differential.
We refer the interested readers to \citet{radiosity}'s article for a more complete introduction.

Radiosity and our method are both closely related to point-based global illumination~\cite{bunnell2005dynamic,buchholz2012quantized,wang2013factorized,wang2015wavelet} that was extensively used in film industry~\cite{christensen2008point} and more recently game industry~\cite{Apers:2024:SDG}. 

\begin{table}[t]
\caption{Common notations and their meanings. In the text, we will also use subscript $i$ to denote quantities belonging to the $i^\text{th}$ Gaussian surfel.}
\centering
\small
\begin{tabular}{ll}
\hline
\textbf{Notation} & \textbf{Meaning} \\ \hline
$\mathbf{a}^d$ & Diffuse albedo \\
$\mathbf{a}^s$ & Specular albedo \\
${B}$ & Radiative flux, or radiosity \\
$\mathbf{B}^c$ & Coefficients for representing radiosity \\
${E}$ & Emission \\
$\mathbf{E}^c$ & Coefficients for representing emission \\
${f}$ & Bidirectional reflectance distribution function \\
$\mathbf{f}^c$ & Coefficients for representing BRDF \\
$\mathcal{G}$ & Unnormalized 2D Gaussian kernel \\
$\mathbf{k}$ & Diffuse specular blending coefficient \\
$L_I(\mathbf{o}, \bm{\omega})$ & Incident radiance given ray origin $\mathbf{o}$ and direction $\bm{\omega}$ \\
$L_O(\mathbf{o}, \bm{\omega})$ & outgoing radiance given location $\mathbf{o}$ and viewing direction $\bm{\omega}$ \\
$N$ & Total number of kernels \\
$\mathbf{n}$ & Normal \\
$P$ & Local tangent plane space \\
$\mathbf{p}$ & Central point \\
$\mathbf{S}=(s^u,s^v)$ & Scaling \\ 
$\mathbf{s}$ & Shininess \\
$\mathbf{T}=(\mathbf{t}^u, \mathbf{t}^v)$ & Principal tangential directions \\
$\mathbf{t}$ & Intersection point between the ray and kernel \\
$V_{ji}$ & Asymmetric fused decay from $j^\text{th}$ kernel to $i^\text{th}$ kernel \\
$Y_{lm}$ & Spherical Harmonic basis function \\
$\alpha(\mathbf{x})$ & Opacity at location $\mathbf{x}$ \\
$\kappa$ & Weighted residual kernel \\
$\Lambda$ & Integrated value of residual kernel for normalization \\
$\bm{\omega}_O$ & outgoing direction (pointing outwards) \\
$\bm{\omega}_I$ & In-coming direction (pointing inwards) \\
\hline
\end{tabular}
\label{table:notation}
\end{table}

We revisit radiosity due to our need for calculating the light transport among discrete primitives. 
Many historical limitations do not exist in our setup or at least can be alleviated using the modern semi-opaque primitives and especially in the context of inverse rendering. 
For example, discontinuity meshing and subdivision are no longer required as we have continuous non-binary visibility.
However, we still face significant challenges. 
The classic radiosity algorithm treats the calculation of geometry relationships (i.e., form factors) as a pre-processing step and focuses on a single forward rendering. 
In our case, we need to calculate the forward rendering at each optimization step which demands an efficient solver. 
The classic radiosity algorithm is also established based on opaque triangle primitives with binary visibility and needs to be extended for Gaussian surfels. 
A differentiable method for radiosity, especially for the backward pass of the radiosity formulation, has also not yet been explored for discrete elements.
We present a method for light transport between Gaussian surfels that address these challenges, by extending radiosity to handle transparency, proposing a Monte-Carlo solver and combine it with existing radiosity solvers~\cite{radiosity_progressive_refinement}, derive efficient formulas for the gradient computation, and present necessary approximations to significantly improve the efficiency while achieving high-quality results.

\section{Preliminaries}
We will first discuss the underlying geometry field and then recap the spherical harmonics basis. See Table~\ref{table:notation} for notations used throughout the paper.

\subsection{Geometry Field}
\label{method-preliminary}
\citet{gfsgs} propose to use an unordered set of $N$ Gaussian surfels to parametrize a stochastic geometry field \cite{object_as_volumes} and then convert it into the density field for differentiable rendering \cite{2DGS}.

\begin{figure}[t]
    \centering
    \includegraphics[width=\linewidth]{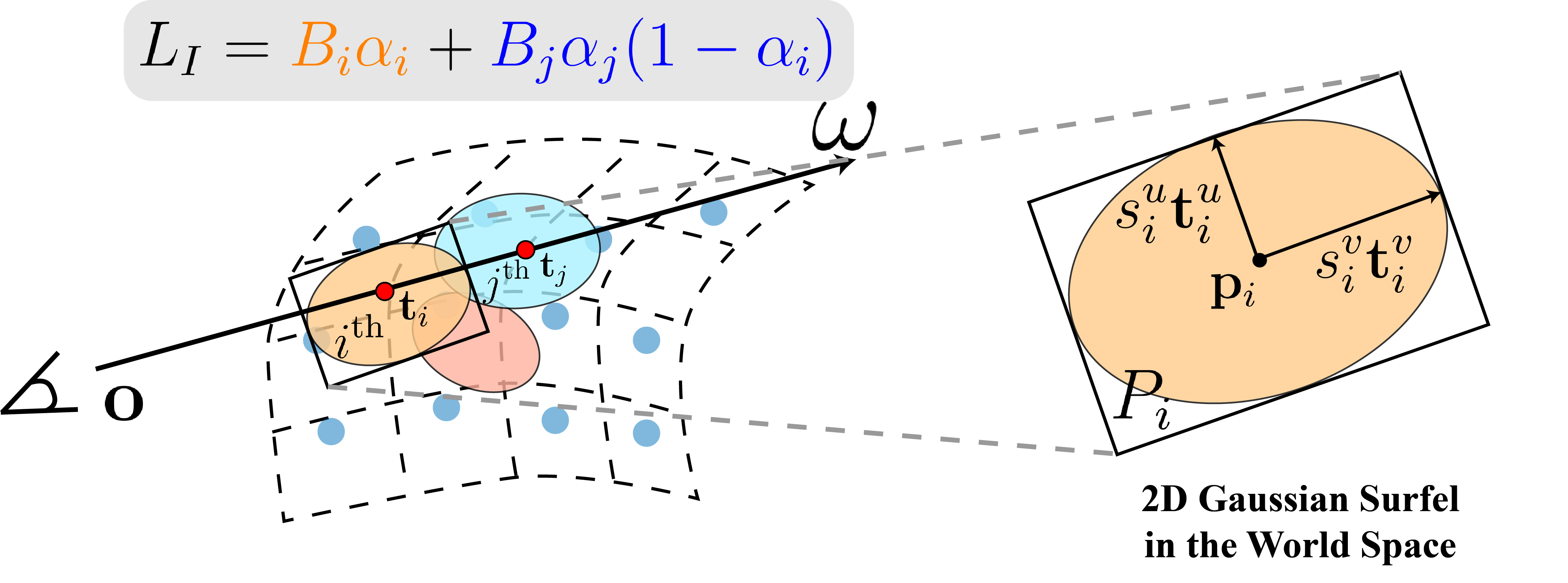}
    \caption{Illustration of the splatting algorithm for rendering when the ray intersects with two Gaussian surfels, i.e., the $i^\text{th}$ and $j^\text{th}$ Gaussian surfels. We also illustrate the parametrization of a 2D Gaussian surfel in the world space.}
    \label{fig:2d-Gaussian}
\end{figure}

Specifically, the $i^\text{th}$ Gaussian surfel is parametrized by a central point $\mathbf{p}_i$, principal tangential directions $\mathbf{T}_i=(\mathbf{t}^u_i, \mathbf{t}^v_i)$, and the scaling in the tangential directions $\mathbf{S}_i=(s^u_i, s^v_i)$, while the normal $\mathbf{n}_i$ can be derived as the cross product of two principal tangential directions. Specifically, these properties define a local tangent plane space $P_i$ and an unnormalized 2D Gaussian kernel (known as \emph{Gaussian surfel}) $\mathcal{G}_i$ illustrated in Fig.~\ref{fig:2d-Gaussian} as:
\begin{equation}
\label{eqn:G}
\begin{aligned}
    P_i(u, v) &= \mathbf{p}_i+s^u_i\mathbf{t}^u_i u + s^v_i \mathbf{t}^v_i v \\
    \mathcal{G}_i(u,v) &= \exp(-\frac{1}{2}(u^2+v^2)), 
\end{aligned}
\end{equation}
where $u, v$ denote the coordinates on the local tangent plane. %

\citet{gfsgs} compose a stochastic geometry field using the defined Gaussian surfels {each of which is further associated with a \emph{geometry value}} and then convert it into the density field for rendering with a refined splatting algorithm. Notably, each Gaussian surfel is associated with a directly learnable outgoing radiance, and all Gaussian surfels are \emph{ideally} sorted based on their intersection depth with the ray. In the volume rendering setting, the incident radiance is then calculated as:
\begin{equation}
\label{eqn:L_I}
    L_I(\mathbf{o}, \bm{\omega}) = \sum_{i=1}^{N} B_i(\mathbf{t}_i, -\bm{\omega}) \alpha_i(\mathbf{t}_i) \prod_{j=1}^{i-1} (1-\alpha_j(\mathbf{t}_j)), 
\end{equation}
where $L_I(\mathbf{o}, \bm{\omega})$ denotes the incident radiance for ray with origin $\mathbf{o}$ and direction $\bm{\omega}$ shooting from the camera plane, $\mathbf{t}_i$ denotes the intersection point between the $i^\text{th}$ kernel and the current ray, $B_i(\mathbf{t}_i, -\bm{\omega})$ denotes the outgoing radiance of the $i^\text{th}$ kernel at $\mathbf{t}_i$ with viewing direction $-\bm{\omega}$, and $\alpha_i(\mathbf{t}_i)$ denotes the calculated footprint value or equivalently, opacity, at $\mathbf{t}_i$. Such a rendering formulation is exact except for the sorting approximation. 
An illustration of the splatting algorithm of Eqn.~\eqref{eqn:L_I} and the parametrization of 2D Gaussian surfels is given in Fig.~\ref{fig:2d-Gaussian}.
Please refer to the paper \cite{gfsgs} for more details. 

\citet{gfsgs} then address the problem of geometry reconstruction from a set of images. 
Given the training views, the corresponding rendered images at those camera positions are supervised using the following loss:
\begin{equation}
\mathcal{L}=\mathcal{L}_\text{rgb}+\lambda_1\mathcal{L}_\text{d}+\lambda_2\mathcal{L}_\text{n}, 
\end{equation}
where $\mathcal{L}_\text{rgb}$ denotes the view synthesis loss \cite{3DGS}, $\mathcal{L}_\text{d}$ denotes the depth distortion loss \cite{2DGS}, $\mathcal{L}_\text{n}$ denotes the depth-normal consistency loss \cite{2DGS,Gaussianshader,normal-depth} and $\lambda_1\text{ is set specific to the dataset},\lambda_2=0.05$. 

In this paper, we address similar tasks, i.e., given a set of images with light sources known or unknown, we achieve relighting and geometry reconstruction. Different from \citet{gfsgs}, we deduce the outgoing radiance or radiosity from the underlying geometry, materials and lighting by differentiable light transport, instead of directly formulating the outgoing radiance as learnable variables.

\begin{figure*}[t]
    \centering
    \includegraphics[width=\linewidth]{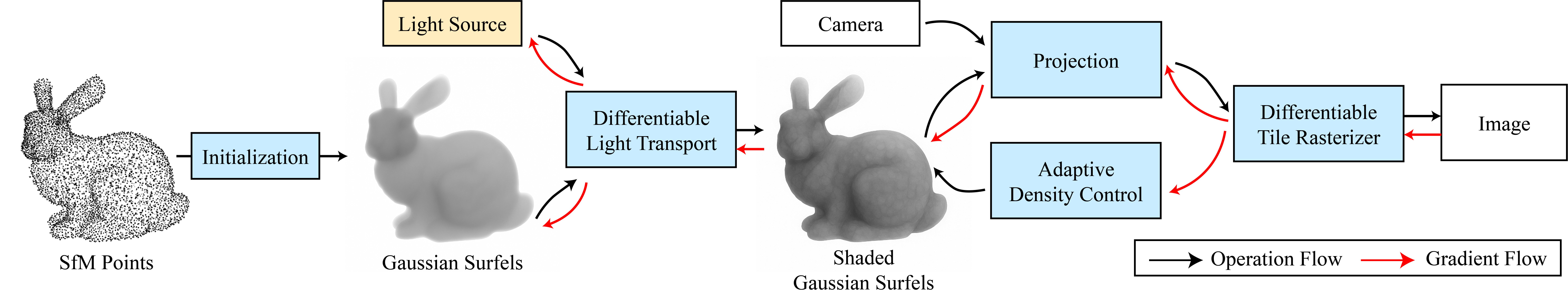}
    \caption{
    \textsc{Overview of our pipeline}. As in previous work~\cite{3DGS,2DGS,gfsgs}, we create a set of 2D Gaussian surfels from a sparse set of Structure-from-Motion (SfM) points if available or random points and then perform the optimization and adaptive density control using image-space supervision. Differently, we incorporate a \emph{differentiable light transport} module to determine the outgoing radiance of each Gaussian surfel from the underlying geometry, material and light sources.}
    \label{fig:pipeline}
\end{figure*}

\subsection{Spherical Harmonics Representation}
\label{method:sh}
As in previous work~\cite{radiosity_sh}, we intend to efficiently store spherical functions, including BRDF, radiance and emission, for each Gaussian surfel, and a popular choice is to use the spherical harmonics basis.
The spherical harmonics basis is introduced into the graphics community to efficiently represent the common spherical rendering-related functions \cite{cabral,radiosity_sh,westin92,sh,sloan_prt}.

We denote the set of real spherical harmonic basis as:
\begin{equation}
    \mathbf{Y}(\bm{\omega})=[Y_{00}(\bm{\omega}),Y_{1,-1}(\bm{\omega}),Y_{10}(\bm{\omega}),Y_{11}(\bm{\omega}),...,Y_{LL}(\bm{\omega})]^T, 
\end{equation}
which is up to degree $L$, and where $\bm{\omega}$ denotes the evaluation direction. We further define $\overline{\mathbf{Y}}(\bm{\omega})=\mathbf{Y}(-\bm{\omega})$.

For any real spherical function $h(\bm{\omega})$, it can be represented as a linear weighted sum of spherical harmonics basis functions:
\begin{equation}
\begin{aligned}
    h(\bm{\omega})&\approx\sum_{l=0}^{L}\sum_{m=-l}^{l}c_{lm}Y_{lm}(\bm{\omega}) \\
    &=\underbrace{[c_{00}, c_{1,-1}, c_{10}, c_{11},...,c_{LL}]}_{\text{coefficient vector}}\mathbf{Y}(\bm{\omega}), 
\end{aligned}
\end{equation}
where we have:
\begin{equation}
\label{sh-coefficient}
    c_{lm} = \int_{\mathcal{S}^2} h(\bm{\omega}) Y_{lm}(\bm{\omega}) d\bm{\omega}.
\end{equation}

We could also expand a 4D real spherical function, i.e., the {reciprocal} BRDF $f$, using the spherical harmonics basis. Specifically, we could view it as a function mapping from a specific incoming direction $\bm{\omega}_I$ into all outgoing directions or equivalently mapping from $\bm{\omega}_I$ to a coefficient vector. Therefore, the evaluation of the BRDF can be defined as:
\begin{equation}
    f(\bm{\omega}_I, \bm{\omega}_O) = \mathbf{f}^T(\bm{\omega}_I) \mathbf{Y}(\bm{\omega}_O), 
\end{equation}
where $\bm{\omega}_O$ denotes the viewing direction, and $\mathbf{f}$ denotes the function mapping from $\bm{\omega}_I$ into a coefficient vector.

As discussed by \citet{radiosity_sh}, a BRDF $f(\bm{\omega}_I, \bm{\omega}_O)$ which is not cosine-weighted and only a function of $\bm{\omega}_I\cdot\bm{\omega}_O$ can be expanded using spherical harmonics as:
\begin{equation}
\label{sh-brdf}
    f(\bm{\omega}_I, \bm{\omega}_O) = \sum_{l=0}^{L}\sum_{m=-l}^{l} c_{lm} \overline{{Y}}_{lm}(\bm{\omega}_I)Y_{lm}(\bm{\omega}_O), 
\end{equation}
where $c_{lm}$ denotes the coefficient for constructing the BRDF and specifically, $c_{lm}=(-1)^{m}c_{l0}$. Therefore, by combining Eqn.~\eqref{sh-coefficient} and Eqn.~\eqref{sh-brdf}, we have:
\begin{equation}
\begin{aligned}
    \mathbf{f}^c &= [c_{00}, c_{1,-1}, c_{01},c_{11},...,c_{LL}] \\
    \mathbf{f}(\bm{\omega}_I) &= \mathbf{f}^c\times \overline{\mathbf{Y}}(\bm{\omega}_I), 
\end{aligned}
\end{equation}
where $\mathbf{f}^c$ is the coefficient vector that defines the BRDF.

Generally, expanding the BRDF into the spherical harmonics basis is non-trivial. However, there is a closed-form solution \cite{sh} for the Phong BRDF \cite{phong} function.

Notably, as discussed by \citet{in-out} and \citet{sh}, the Phong BRDF also approximates the Blinn Phong BRDF~\cite{blinn_phong} and microfacet BRDFs, such as the Torrance-Sparrow model~\cite{Torrance}, when it is expanded in the spherical harmonics basis space. 
\citet{in-out} show that the supported maximum accurate shininess of the Phong BRDF using up to $L$ degree spherical harmonics is $\sim\frac{1}{5}(L+1)^2$.

\section{Overview}

Our goal is to compute a view-dependent radiance field that consists of Gaussian surfels and spherical harmonic coefficients living on them, while accounting for light transport between the surfels. Instead of path tracing, which could be noisy and slow in our case, we revisit classical radiosity, extend it to handle transparency, and derive its differential counterpart. We found that this offers several benefits: 1) it significantly simplifies some derivative computations and 2) it allows us to adapt radiosity solvers that cache quantities within surfels.

We address the problem of relighting and geometry reconstruction from a set of training images with known or unknown light sources. During test time, we perform relighting at the designated novel viewpoints and lighting conditions. For geometry reconstruction, we extract the mesh from rendered depth maps of all training views using TSDF fusion \cite{open3d, tsdf}. 

Figure~\ref{fig:pipeline} provides an overview of our pipeline. Specifically, we initialize from structure-from-motion (SfM) points if available or random points, and calculate the outgoing radiance from the underlying geometry, material and lighting conditions, which is denoted as ``Differentiable Light Transport'' and is the core of our method. 
After the Gaussian surfels receive their calculated outgoing radiance, they are then \emph{shaded} and fed into the differentiable tile-based rasterizer for efficient rendering.

We use exactly the same optimization procedure as in \citet{gfsgs}'s work, which is described in Sec.~\ref{method-preliminary}, except that we further incorporate a masking loss $\mathcal{L}_\text{a}$ with weight $0.1$ to supervise the rendered alpha mapping using the L1 norm compared with the ground truth, and an optional per-pixel L2 norm loss $\mathcal{L}_\text{env}$ with weight $0.05$ to regularize the intensity of the optimized environment map (if an environment map is used). 

As in previous work~\cite{2DGS,3DGS, gfsgs}, we also use adaptive density control strategies to grow the primitives.

Regarding the differentiable light transport, we will first discuss our forward rendering framework (Sec.~\ref{sec:forward-rendering}).
Subsequently, we present the gradient calculation for our global illumination framework (Sec.~\ref{method-backward}).
Finally, we discuss the solvers for light transport (Sec.~\ref{method-solver}) and the introduced approximations (Sec.~\ref{method-approx}). 

Due to the different conventions, we use \emph{radiosity}, and \emph{outgoing radiance} interchangeably in the paper as we also model the directional effects. We also use symbols $\times$ and $/$ to denote element-wise multiplication and division of vectors for simplicity. When an operator is omitted, it usually means that the operator is (element-wise) multiplication, except that it refers to matrix multiplication when two operands are a row-vector and a column-vector.

\section{Forward Rendering}
\label{sec:forward-rendering}
We will first discuss our representations of geometry, materials, emission, outgoing radiance, and lighting. Afterwards, we present our framework for forward rendering, including a special case where we only consider the direct illumination, and the general case where we consider the global illumination.

\subsection{Geometry, Material, Emission, Outgoing Radiance, and Lighting Representations}
\label{method-representation}
\paragraph{Geometry.} As discussed in Sec.~\ref{method-preliminary}, the geometry is represented with a stochastic geometry field using the semi-opaque primitives, i.e., Gaussian surfels. The visibility is therefore non-binary. We disable the self-reflection and treat each primitive as a two-sided primitive with both reflective front and non-reflective back faces for simplicity.
\paragraph{Material, Emission and outgoing Radiance.} We now introduce our first {assumption} that 
the material, emission and outgoing radiance are constant within the Gaussian surfel support.
This assumption is also widely adopted in the radiosity theory and the splatting algorithm~\cite{3DGS, 2DGS, volume-splatting, gfsgs}, which naturally suits the discrete representation of geometry. %

Specifically, the $i^\text{th}$ kernel is associated with a {reciprocal} bidirectional reflectance distribution function (BRDF) $f_i$ which is a 4D function that takes incoming direction and outgoing direction as input, and an emission function $E_i$ and an outgoing radiance function $B_i$ which take the outgoing direction as input. %
As discussed in Sec.~\ref{method:sh}, the emission $E_i$ and outgoing radiance $B_i$ can be represented as spherical harmonics:
\begin{equation}
\begin{aligned}
    E_i(\bm{\omega}) &= (\mathbf{E}^c_i)^T \mathbf{Y}(\bm{\omega}) = \mathbf{Y}^T(\bm{\omega}) \mathbf{E}^c_i \\
    B_i(\bm{\omega}) &= (\mathbf{B}^c_i)^T \mathbf{Y}(\bm{\omega}) = \mathbf{Y}^T(\bm{\omega}) \mathbf{B}^c_i, 
\end{aligned}
\end{equation}
where $\mathbf{E}^c_i$ and $\mathbf{B}^c_i$ denote the coefficient vectors for emission and radiosity.

We use a mixture of diffuse and Phong models to represent the BRDF. For each $i^\text{th}$ kernel, it is associated with diffuse albedo $\mathbf{a}^d_i$, specular albedo $\mathbf{a}^s_i$, shininess $\mathbf{s}_i$, and diffuse specular blending coefficient $\mathbf{k}_i$. 
The BRDF $f_i$ is defined as:
\begin{equation}
    \label{eqn:brdf}
        f_i(\bm{\omega}_I, \bm{\omega}_O) = \mathbf{k}_i \frac{\mathbf{a}^d_i}{\pi} + (1-\mathbf{k}_i) \mathbf{a}^s_i \frac{\mathbf{s}_i+1}{2\pi}|\bm{\omega}_r\cdot\bm{\omega}_I|^\mathbf{s_i}, 
    \end{equation}
where $\bm{\omega}_r$ denotes the reflected outgoing direction with respect to the surface normal. 
Please refer to Appendix~\ref{app:brdf} for the formula of the coefficient vector $\mathbf{f}^c_i$ corresponding to $f_i$, using the conclusions in \citet{sh}'s work.

\paragraph{Lighting.} We represent all light sources as point light sources in the form of special infinitesimal emissive Gaussian surfels for simplicity. %
We also support directional light sources approximately, by putting the point light sources far away and disabling the inverse square falloff from the light sources to the kernels. 
{Our method can also handle environment maps in a similar way.
Since we derive the derivatives with respect to the emission and position below, our method supports the optimization of lighting as well.}

\subsection{Direct Illumination}
\label{method-direct}
We first consider a special case where only direct illumination is taken into account. For each Gaussian surfel, we only consider the incident radiance from the light sources, and the final outgoing radiance is a weighted average of the response at all the points within the local support $P$ of the kernel. This weighted averaging is described as a kernel $\kappa$, such as Gaussian, Uniform, Dirac delta, \emph{etc.}, with support $P$, and we define $\Lambda=\int_P\kappa(\mathbf{x})d\mathbf{x}$.

The detailed derivation for direct lighting can be seen as a special case of the framework for global illumination that we will discuss later. In conclusion, we have the following rendering equation:
\titledeq{\xspace forward rendering equation (direct illumination)}{primalForeground}{primalBackground}{
	\begin{equation}
        \begin{aligned}
            \mathbf{B}^c_i =& \mathbf{E}^c_i + \frac{\alpha_i(\mathbf{p_i})}{\Lambda_i}\times \\
            &\sum_{j\in\text{LS}}\int_{P_i}
        \mathcalloutbox{BRDF}{gray!15}{gray}{\mathbf{f}_i(\bm{\omega}_I)}\times 
        \mathcalloutbox{Emission}{violet!15}{violet}{(\mathbf{Y}(\bm{\omega}_I)^T\mathbf{E}^c_j)}\times 
        \mathcalloutbox{Decay}{cyan!15}{cyan}{{V}_{ji}(\mathbf{p}_j, \mathbf{x})} d\mathbf{x},\\
            &\text{where LS denotes the set of light sources.}
        \end{aligned}
        \end{equation}
}
In this equation, $\alpha_i(\mathbf{p}_i)$ is needed as the semi-transparent kernel does not absorb all the energy, and we have:
\begin{equation}
\begin{aligned}
    \bm{\omega}_I&=\frac{\mathbf{x}-\mathbf{p}_j}{||\mathbf{x}-\mathbf{p}_j||_2^2} \\
    \mathbf{f}_i(\bm{\omega}_I)&=\mathbf{f}^c_i\times \overline{\mathbf{Y}}(\bm{\omega}_I) \\
    V_{ji}(\mathbf{p}_j, \mathbf{x})&=\frac{|\mathbf{n}_i\cdot\bm{\omega}_I|}{||\mathbf{x}-\mathbf{p}_j||_2^2}\kappa_i(\mathbf{x})\prod_{k=1}^{j-1}(1-\alpha_k), \\
\end{aligned}
\end{equation}
where $\mathbf{B}^c_i,\mathbf{E}^c_i,\mathbf{f}^c_i$ denote the coefficients for constructing the radiosity, emission, and BRDF using spherical harmonics. $V_{ji}$ represents the fused decay component as it includes the angular factor, averaging kernel and visibility that is the standard form for Gaussian splatting, radiance field, or volumetric rendering.

Notice that this equation is not recursive and can be directly evaluated. The gradient calculation can also be efficiently solved by the auto-differentiation.

\paragraph{Discussion} As shown in Fig.~\ref{fig:teaser}, partial understanding of full light transport, or direct illumination alone can already provide benefits for relatively sparse-view relighting, compared to data-driven approaches. Compared to standard direct lighting in ray tracing, the direct illumination in our case effectively caches the radiance within each kernel, enabling view-independent rendering as with the classical radiosity theory. 
Compared to these previous works of relighting with far-field light sources using Gaussian splatting (e.g., \cite{Gaussianshader, relightable3DGS, GIGS, GSIR}), we do not ignore, freeze or binarize the visibility, support both near-field and far-field light sources and derive our rendering equation from physically-based rendering to ensure the correctness.

\subsection{Global Illumination}
\label{method-forward}
\begin{figure}
    \centering
    \includegraphics[width=\linewidth]{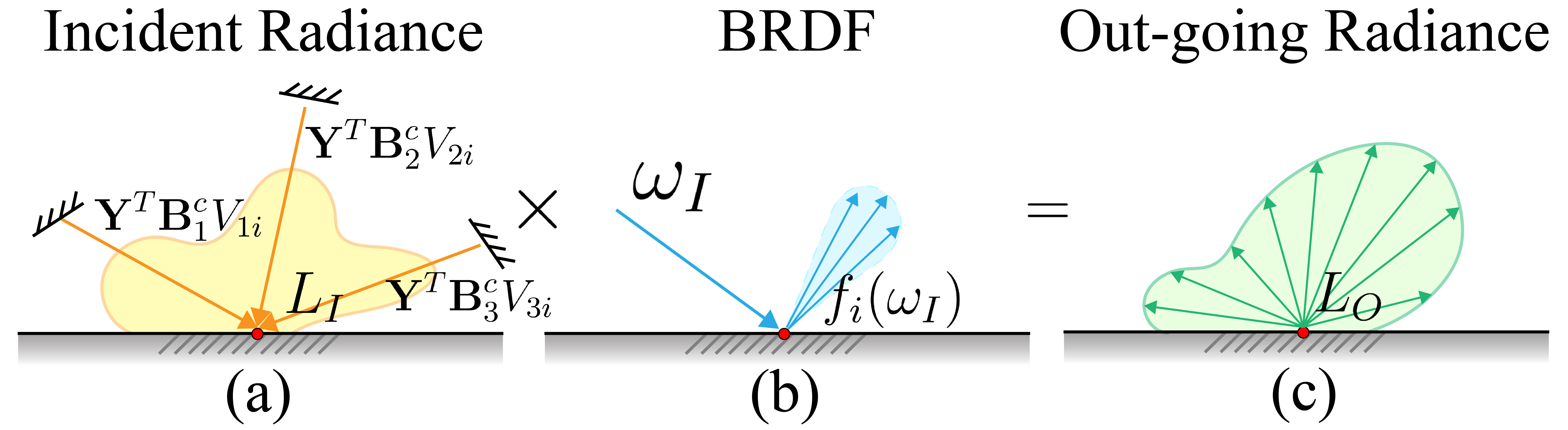}
    \caption{Illustration of calculating the outgoing radiance for the $i^\text{th}$ kernel in Eqn.~\eqref{eqn:final-radiosity}. We omit the emission for simplicity. The $i^\text{th}$ kernel first receives radiance from all other kernels in space as the incident radiance, which is then converted to the outgoing radiance through the underlying material, i.e., BRDF.}
    \label{fig:li_brdf_lo}
\end{figure}

We now present our forward rendering model for global illumination. Its derivation comes from the rendering equation~\cite{rendering_equation} and relies on an assumption that any two kernels are non-overlapping to solve the established forward rendering system using the weighted residual form. %
See Appendix~\ref{app:forward} for more details. 

We cannot directly borrow the classical radiosity equation due to our change from fully opaque surface primitives into semi-opaque primitives, even though the eventual equation indeed mimics the radiosity equation with small but important changes.

The forward rendering equation is given as:
\begin{equation}
    \label{eqn:before-final-radiosity}
    \begin{aligned}
        \mathbf{B}^c_i =& {\mathbf{E}^c_i}+\frac{\alpha_i(\mathbf{p}_i)}{\Lambda_i}\times\\
        &\sum_{j=1}^{N}\int_{P_i}\int_{P_j}
        \mathcalloutbox{BRDF}{gray!15}{gray}{\mathbf{f}_i(\bm{\omega}_I)}\times 
        \mathcalloutbox{Radiance}{orange!15}{orange}{(\mathbf{Y}(\bm{\omega}_I)^T\mathbf{B}^c_j)}\times 
        \mathcalloutbox{Decay}{cyan!15}{cyan}{{V}_{ji}(\mathbf{x}', \mathbf{x})} d\mathbf{x}' d\mathbf{x}, 
    \end{aligned}
\end{equation}
where we have:
\begin{equation}
\begin{aligned}
    \bm{\omega}_I&=\frac{\mathbf{x}-\mathbf{x}'}{||\mathbf{x}-\mathbf{x}'||_2^2} \\
    V_{ji}(\mathbf{x}', \mathbf{x})&=\frac{|\mathbf{n}_i\cdot\bm{\omega}_I||\mathbf{n}_j\cdot\bm{\omega}_I|}{||\mathbf{x}-\mathbf{x'}||_2^2}\kappa_i(\mathbf{x})\alpha_j(\mathbf{x}') \prod_{k=1}^{j-1}(1-\alpha_k), 
\end{aligned}
\end{equation}
and the meanings of other quantities, such as $\mathbf{f}_i(\bm{\omega}_I), \mathbf{B}^c_i,\mathbf{E}^c_i,\mathbf{f}^c_i$ follow those in the direct illumination case (Sec.~\ref{method-direct}).

As discussed before, we treat point light sources as special Gaussian surfels to simplify the framework. Specifically, when the selected $j^\text{th}$ kernel represents a light source, we do not integrate over the $P_j$ but directly sample its central point as $\mathbf{x}'$ and disable the dot product between the normal and incoming direction for the $j^\text{th}$ kernel in $V_{ji}$. When it represents the directional light, we further disable the inverse square falloff in $V_{ji}$. Note that by only selecting $j$ within the set of light sources, we reach the special case where we only consider the direct illumination in Sec.~\ref{method-direct}.

We further absorb the $\alpha_i(\mathbf{p}_i)/\Lambda_i$ term into the $\mathbf{f}_i(\bm{\omega}_I)$ term to simplify the theoretical deduction, and the resulting equation is then:
\titledeq{\xspace forward rendering equation}{primalForeground}{primalBackground}{
	\begin{equation}
        \label{eqn:final-radiosity}
        \begin{aligned}
            \mathbf{B}^c_i &= \mathbf{E}^c_i + \sum_{j=1}^{N}\int_{P_i}\int_{P_j}{\mathbf{f}_i(\bm{\omega}_I)} {(\mathbf{Y}(\bm{\omega}_I)^T\mathbf{B}^c_j)} {V}_{ji} d\mathbf{x}' d\mathbf{x}  
        \end{aligned}
        \end{equation}
}
This equation resembles the classical radiosity equation \cite{radiosity_sh} after absorbing the terms, but the meaning of symbols (i.e., $\mathbf{f}_i$, $V_{ji}$) differs and the visibility is non-binary in our case. 
Notice that this equation is recursive as $\mathbf{B}^c$ shows up in both sides of the equation, and we will discuss how to solve it later.
We provide a visualization of this equation in Fig.~\ref{fig:li_brdf_lo}.

To summarize, we parametrize the geometry and materials using Gaussian surfels and derive the analytical formula from the rendering equation for the outgoing radiance. We make the assumptions that: 1) The material, emission and outgoing radiance are constant within the kernel; 2) There is no overlapping between any two Gaussian surfels. After we calculate the outgoing radiance for each surfel, we could render the image given an arbitrary viewpoint using Eqn.~\eqref{eqn:L_I}, which is implemented as an efficient rasterization process~\cite{2DGS}.

\section{Gradients for Global Illumination}
\label{method-backward}
For optimization, we need to deduce the partial derivative of the loss function $\mathcal{L}$ with respect to the underlying geometry, materials, and lighting conditions from the partial derivative of the loss function with respect to the calculated outgoing radiance, i.e., $\partial\mathcal{L}/\partial \mathbf{B}^c_i$, which comes from the differentiable tile-based rasterizer.

Compared to using automatic differentiation with techniques such as path replay backpropagation~\cite{PathReplay}, we show that, by the explicit derivation of gradients, there is a \emph{duality} between the forward pass and backward pass, which further allows non-recursive analytical definitions of gradients, which are accurate and efficient to calculate.
The detailed derivation is given in Appendix~\ref{app:backward}, with the main results presented here.

\paragraph{Gradients with respect to the emissions.} We first consider the partial derivatives for the emissions, as they will be used in calculating other gradients as well. 
The $\partial\mathcal{L}/\partial\mathbf{E}^c_i$ are given as:
\titledeq{\xspace gradients with respect to emissions}{differentialForeground}{differentialBackground}{
	\begin{equation}
        \label{eqn:final_dL_dE}
        \begin{aligned}
            \frac{\partial\mathcal{L}}{\partial\mathbf{E}^c_i} =& \frac{\partial\mathcal{L}}{\partial\mathbf{B}^c_i} + \frac{1}{\mathbf{f}^c_i} \sum_{j=1}^{N}\int_{P_i}\int_{P_{j}}{\mathbf{f}_i}(\bm{\omega}_I) (\mathbf{Y}(\bm{\omega}_I)^T({\mathbf{f}^c_j}\times \frac{\partial\mathcal{L}}{\partial\mathbf{E}^c_j})) \overline{{V}_{ji}} d\mathbf{x}' d\mathbf{x}, \\
            &\text{where }\overline{V_{ji}}=V_{ij}.
        \end{aligned}
        \end{equation}
}
Note that this equation closely resembles Eqn.~\eqref{eqn:final-radiosity}, except that the decay term is reversed and we are propagating the gradients instead of the emission. This relationship implies the duality between forward rendering and gradient calculation, which is analogous to previous inverse and differentiable rendering works~\cite{duality,adjoint_differentiable}. We will further extend this duality by deducing the analytical non-recursive formulae for materials and geometry. Therefore, the solvers we develop later will generalize to both forward rendering and gradient calculation. %

\paragraph{Gradients with respect to the BRDF}
We now consider the partial derivatives with respect to the BRDF, or specifically the coefficients $\mathbf{f}^c$. Perhaps surprisingly, $\partial\mathcal{L}/\partial\mathbf{f}^c$ can be analytically solved using the previously calculated $\mathbf{B}^c$ and $\partial\mathcal{L}/\partial\mathbf{E}^c$ by noticing that $\mathbf{B}^c$ represents the summation of all possible propagated emissions and $\partial\mathcal{L}/\partial\mathbf{E}^c$ represents the summation of all possible propagated gradients. The full derivation is given in Appendix~\ref{app:backward}.

Specifically, we have:
\titledeq{\xspace gradients with respect to materials}{differentialForeground}{differentialBackground}{
	\begin{equation}
        \label{eqn:final-dl-dbrdf}
        \begin{aligned}
            \frac{\partial\mathcal{L}}{\partial\mathbf{f}^c_i} =\frac{\partial\mathcal{L}}{\partial\mathbf{E}^c_i}\times \frac{(\mathbf{B}^c_i-\mathbf{E}^c_i)}{\mathbf{f}^c_i} 
        \end{aligned}
        \end{equation}
}
This equation is an analytical non-recursive accurate closed-form solution, which is therefore more efficient than using standard automatic differentiation.

\paragraph{Gradients with respect to the geometry}
The remaining gradients are the partial derivatives with respect to the geometric properties (i.e., central points, scaling, \emph{etc.}) from the decay term and directional effects. The derivation for their analytical non-recursive forms is similar to the derivation of $\partial\mathcal{L}/\partial\mathbf{f}^c_i$, which is elaborated in Appendix~\ref{app:backward}.

Without losing generality, we define $\beta_i$ as a placeholder for the geometric property of the $i^\text{th}$ kernel, such as the $\mathbf{p}_i,\mathbf{S}_i,\mathbf{T}_i,etc.$, that we want to differentiate through.

Specifically, from the decay term $V$, the first part of $\partial\mathcal{L}/\partial \beta_i$ is given as:
\titledeq{\xspace gradients w.r.t. geometry from the decay term}{differentialForeground}{differentialBackground}{
	\begin{equation}
        \label{eqn:dl-dgeo-0}
        \begin{aligned}
            \frac{\partial\mathcal{L}}{\partial \beta_i} =& \sum_{a=1}^N \sum_{b=1}^N \int_{P_b}\int_{P_a} \frac{(\overline{\mathbf{Y}}^T({\mathbf{f}^c_b}\times\frac{\partial\mathcal{L}}{\partial\mathbf{E}^c_b}))(\mathbf{Y}^T\mathbf{B}^c_a)}{s^u_as^v_as^u_bs^v_b} \cdot \\
            & \frac{\partial({V}_{ab}s^u_as^v_as^u_bs^v_b)}{\partial \beta_i}  d\mathbf{x}'d\mathbf{x}
        \end{aligned}    
        \end{equation}
}
This formula for gradients is again analytical and non-recursive, but we need to enumerate every pair of kernels as $a$ and $b$ in Eqn.~\eqref{eqn:dl-dgeo-0}, which is costly. In practice, since we should solve the light transport in the forward pass, we record each pair of kernels that are visible to each other and only calculate the gradients from these pairs since the decay term is non-zero.

Besides, we also need to take the directional effects into account. The second and last part of $\frac{\partial\mathcal{L}}{\partial \beta_i}$ is given as:
\titledeq{\xspace gradients w.r.t. geometry from directional effects}{differentialForeground}{differentialBackground}{
	\begin{equation}
        \label{eqn:dl-dgeo-1}
        \begin{aligned}
            \frac{\partial\mathcal{L}}{\partial \beta_i} &= \sum_{j=1}^N \int_{P_i}\int_{P_j} [{V}_{ji} (\mathbf{Y}^T \mathbf{B}^c_j) (\mathbf{f}^c_i\times\frac{\partial\mathcal{L}}{\partial\mathbf{E}^c_i})]^T \overline{\mathbf{Y}}'(\bm{\omega}_I) \frac{\partial\bm{\omega}_I}{\partial \beta_i} d\mathbf{x}' d\mathbf{x} \\
            &+ \sum_{j=1}^N \int_{P_j}\int_{P_i} [\overline{{V}_{ji}} (\overline{\mathbf{Y}}^T({\mathbf{f}^c_j}\times \frac{\partial\mathcal{L}}{\partial\mathbf{E}^c_j})) \mathbf{B}^c_i]^T \mathbf{Y}'(\bm{\omega}_I) \frac{\partial\bm{\omega}_I}{\partial \beta_i} d\mathbf{x}' d\mathbf{x} 
        \end{aligned}    
        \end{equation}
}
Similarly, this formula is also analytical and non-recursive and we only calculate the gradients from kernels that are visible to the $i^\text{th}$ kernel.

In conclusion, the gradient calculation in our case is different from that in path tracing. 
We first define the gradients with respect to the emissions in Eqn.~\eqref{eqn:final_dL_dE}, which is shown to be equivalent to a forward pass and also observed by \citet{adjoint_differentiable}. Different from path tracing gradient calculation, we observe that the emission derivatives are used in many other derivatives, which leads to analytical non-recursive definitions for other derivatives. Moreover, the solvers we design later will cache the necessary quantities within the surfels, preventing redundant computation during the backward pass.

We conduct an ablation study in Sec.~\ref{sec-exp-ablation} to compare standard auto-differentiation method with our proposed, more efficient gradient computation method.
\section{Solvers for Global Illumination}
\label{method-solver}
After establishing the light transport framework, we develop solvers to solve the recursive formulation in Eqn.~\eqref{eqn:final-radiosity} or Eqn.~\eqref{eqn:final_dL_dE}. 
Typically, in the classical radiosity theory, the recursive formulation is transformed into a linear system, and the exact solutions are given by solving the linear system. However, maintaining the whole linear system {using methods such as extended form factors} and computing the exact solutions is {computationally} infeasible in our case since we need to solve the radiosity system at each optimization step where geometry, materials, \emph{etc.} keep changing. We will first discuss the existing solver, i.e., progressive refinement \cite{radiosity_progressive_refinement}, and then introduce our Monte-Carlo solver which efficiently finds the solutions in an unbiased manner and finally discuss how to hybridize these two solvers to improve the efficiency and reduce the variance. %
Due to the duality of the forward pass and gradient calculation, we will use the forward pass as the illustrative example here without losing generality.

\paragraph{Progressive Refinement.} \citet{radiosity_progressive_refinement} propose the progressive refinement as an iterative solver from the shooting perspective for the purpose of interactive rendering. It first initializes the radiosity as the emission, and iteratively selects a kernel and shoots its unshot radiance into all other kernels. %
We first transform Eqn.~\eqref{eqn:final-radiosity} into the shooting perspective, i.e., the $i^\text{th}$ kernel shoots its radiance into all other $j^\text{th}$ kernels, as:
\titledeq{\xspace forward rendering equation (shooting perspective)}{primalForeground}{primalBackground}{
	\begin{equation}
            \forall j\in\{1,2,...,N\}, \mathbf{B}^c_j \leftarrow \mathbf{B}^c_j + \int_{P_j} \int_{P_i} \mathbf{f}_j (\mathbf{Y}^T\mathbf{B}^c_i) {V}_{ij} d\mathbf{x}'d\mathbf{x} 
        \end{equation}
}

We could apply exactly the same progressive refinement algorithm from \citet{radiosity_progressive_refinement}. Specifically, we initialize $\mathbf{B}^c_j$ as $\mathbf{E}^c_j$. %
At each step, we choose a shooting index %
and then calculate how much it emits to all other kernels %
and update their radiance. %
If we do not use progressive refinement to solve the full solution, we will get intermediate calculated radiance $\widehat{\mathbf{B}^c_i}$ and remaining unshot radiance $\delta\mathbf{B}^c_i$ for the $i^\text{th}$ kernel.

The selection of the shooting index %
is usually heuristic. For example, we could select the index that has the largest intensity of unshot outgoing radiance. This corresponds to the sorted order in the original progressive refinement method. %

However, simply applying progressive refinement cannot efficiently reach the final converged solution as we can only select limited shooting indices at a time while we could have hundreds of thousands of Gaussian kernels. We observe that there is a nice property that the solving of remaining radiance given unshot outgoing radiance is a recursive formula in the same form as Eqn.~\eqref{eqn:final-radiosity}, i.e.,
\begin{equation}
\begin{aligned}
    \mathbf{B}^r_i =& \delta\mathbf{B}^c_i + 
\sum_{j=1}^{N}\int_{P_i}\int_{P_j}{\mathbf{f}_i} {(\mathbf{Y}^T\mathbf{B}^r_j)} {V}_{ji} d\mathbf{x}' d\mathbf{x}, \\
\end{aligned}
\end{equation}
where $\mathbf{B}^r_j$ denotes the remaining radiance in the coefficient space of the spherical harmonics basis for the $i^\text{th}$ kernel.
And we have the final outgoing radiance as:
\begin{equation}
    \mathbf{B}^c_i = \widehat{\mathbf{B}^c_i} + \mathbf{B}^r_i - \delta\mathbf{B}^c_i.
\end{equation}

This property allows us to hybridize the progressive refinement with other solvers, or specifically the Monte-Carlo solver that we will discuss next, to solve the light transport.

\paragraph{Monte-Carlo Solver.} From Eqn.~\eqref{eqn:final-radiosity}, we now apply a one-point Monte-Carlo estimator to address the summation in the equation:
\begin{equation}
\label{eqn:one-point-estimator}
    \mathbf{B}^c_i \approx \mathbf{E}^c_i + \frac{1}{\Pr[\text{choose }j]}\int_{P_i}\int_{P_j}{\mathbf{f}_i} {(\mathbf{Y}^T\mathbf{B}^c_j)}{V}_{ji} d\mathbf{x}' d\mathbf{x}, 
\end{equation}
where the choice of $j$ will be discussed later. The familiar reader will recognize that it looks like a standard path-tracing formulation. In path tracing, the Monte-Carlo solution is achieved by iteratively selecting the next direction, applying Russian roulette, terminating at the preset maximum depth, \emph{etc}. In contrast, instead of iteratively selecting the next direction, we iteratively select the next kernel as the next event. Due to the discretizations, we are explicitly solving and \emph{caching} the outgoing radiance for all kernels simultaneously, which enables us to apply a technique known as \emph{off-policy temporal difference learning} (known as TD(0)) \cite{reinforcement} in the reinforcement domain. %

\begin{algorithm}[t]
\caption{Monte-Carlo Solver Algorithm}
\label{monte-carlo-solver}
\begin{algorithmic}[1]
\REQUIRE All Gaussian kernels with their properties, emissions and step $T$.
\ENSURE Estimated outgoing radiance.

\FullComment{Initialize sum of all estimations of radiance}
\STATE $\forall i\in\{1,2,\dots,N\},\ \mathbf{S}_i^B \gets 0$ \label{alg:init-start}

\FullComment{Initialize sum of the square of all estimations of radiance}
\STATE $\forall i\in\{1,2,\dots,N\},\ \mathbf{S}_i^{B^2} \gets 0$

\FullComment{Initialize number being visited}
\STATE $\forall i\in\{1,2,\dots,N\},\ D_i \gets 0$ \label{alg:init-end}

\FullComment{Enumerate each step}
\FOR{$t \textbf{ in } 1,2,\dots,T$}
  \FullComment{Select kernels for receiving radiance}
  \IF{$t \le 8$} \label{alg:select-start}
    \FullComment{Select all kernels}
    \STATE $\{a_1,\dots,a_N\} \gets \{1,2,\dots,N\}$
  \ELSE
    \FullComment{Select $N$ kernels based on variance}
    \STATE $\{a_1,\dots,a_N\} \gets {SelectStartingIndex}$ \label{alg:mc-select}
  \ENDIF \label{alg:select-end}

  \FullComment{Stochastically find kernels for emitting radiance}
  \STATE $\{b_1,p_1,\dots,b_N,p_N\} \gets {NextEvent}$ \label{alg:mc-next-event}

  \FullComment{Calculate the new estimation of received radiance}
  \STATE $\forall i\in\{1,2,\dots,N\},\ \delta_i \gets NE\!\left(a_i,b_i,p_i,\ \mathbf{E}^c_{a_i},\ \mathbf{S}_{b_i}^B/D_{b_i}\right)$ \label{alg:mc-ne}

  \FullComment{Update the sum of all estimations of radiance}
  \STATE $\forall i\in\{1,2,\dots,N\},\ \mathbf{S}_{a_i}^B \gets \mathbf{S}^B_{a_i}+\delta_i$ \label{alg:update-start}

  \FullComment{Update the sum of the square of all estimations of radiance}
  \STATE $\forall i\in\{1,2,\dots,N\},\ \mathbf{S}^{B^2}_{a_i} \gets \mathbf{S}^{B^2}_{a_i}+\delta_i^2$

  \FullComment{Update the number of being visited}
  \STATE $\forall i\in\{1,2,\dots,N\},\ D_{a_i} \gets D_{a_i}+1$ \label{alg:update-end}
\ENDFOR

\FullComment{Calculate the radiance as the average of all estimations}
\STATE $\forall i\in\{1,2,\dots,N\},\ \mathbf{B}^c_i \gets \mathbf{S}^B_i/D_i$
\STATE \textbf{return} $\{\mathbf{B}^c_1,\dots,\mathbf{B}^c_N\}$
\end{algorithmic}
\end{algorithm}

We summarize the Monte-Carlo solver in Alg.~\ref{monte-carlo-solver}. 
Specifically, it is still an iterative procedure with time steps $0,1,...,T-1$, where $T$ denotes the maximum time step. At each time $t\in\{0,1,...,T-1\}$, we maintain a version of the estimated outgoing radiance for every kernel, which is denoted as ${\mathbf{B}^c_i}^{(t)}$. Starting from an initial guess that $\forall i\in\{1,2,...,N\}, {\mathbf{B}^c_i}^{(0)}=\mathbf{0}$ (Line~\ref{alg:init-start}-\ref{alg:init-end} in Alg.~\ref{monte-carlo-solver}), the update rule is defined following TD(0) as:
\titledeq{\xspace monte-carlo solver update rule}{primalForeground}{primalBackground}{
	\begin{equation}
        \label{eqn:update}
        \begin{aligned}
            {\mathbf{B}^c_i}^{(t+1)}=&\frac{t}{t+1}\times
            {{\mathbf{B}^c_i}^{(t)}}+\frac{1}{t+1}\times\\
            &(\mathbf{E}^c_i+\frac{1}{\Pr[\text{choose }j]}\int_{P_i}\int_{P_j}{\mathbf{f}_i} {{(\mathbf{Y}^T\mathbf{B}^c_j}^{(t)})} {V}_{ji} d\mathbf{x}' d\mathbf{x}) 
        \end{aligned}
        \end{equation}
}
Effectively, we blend the old estimate of $\mathbf{B}^c_i$ with the new estimate calculated from Eqn.~\eqref{eqn:one-point-estimator} (Line~\ref{alg:mc-ne} in Alg.~\ref{monte-carlo-solver}). It is proven to converge when the learning rate ($\frac{t}{t+1}$ and $\frac{1}{t+1}$ coefficients) satisfies the Robbins-Monro condition~\shortcite{robbin} {and has been used in many-lights rendering} before \cite{learning-to-cluster}.

We update the estimated outgoing radiance simultaneously for all selected $N$ kernels and all directions (Line~\ref{alg:update-start}-\ref{alg:update-end} in Alg.~\ref{monte-carlo-solver}) since we operate on the coefficient space of the spherical harmonics basis. Notice that the update rule is implemented as an average due to the usage of a geometric series as the learning rate for temporal difference learning.

Our update algorithm is parallel-friendly, and the maximum time step $T$ here is similar to the number of Monte-Carlo samples for a pixel in ray tracing. As the iteration increases, the depth of light transport also grows, which eliminates the need of setting a maximum depth. In practice, we empirically set $T=64$ to achieve a balance between efficiency and variance. This approach is different from \emph{radiance caching}~\cite{radiance_caching} as in that case, the outgoing radiance needs to be \emph{fully independently} solved for the selected points, and then later used as cached values in the final solving procedure. This approach is also different from the iterative solution method for linear systems in the classical radiosity. In the iterative solution method, the intermediate results cannot be used as an unbiased estimation of the final solution, while ours can.

\paragraph{Next event estimation.} We now present the algorithm to choose the $j$ or, formally, estimate the next event (Line~\ref{alg:mc-next-event} in Alg.~\ref{monte-carlo-solver}) for the $i^\text{th}$ kernel. As is well understood, the probability of choosing the kernel $j$, denoted as $p_j$, should be proportional to the integral in Eqn.~\eqref{eqn:one-point-estimator}. The integral can be approximated by removing the BRDF response and the expensive transmission term within the decay term:
\begin{equation}
\label{eqn:importance-sampling}
    p_j \propto \underbrace{\text{clamp}({{\mathbf{B}^c_{j}}^{{(t)}}}^{T}Y)}_{\text{how much emit}} \underbrace{\frac{|\mathbf{n}_i\cdot\bm{\omega}_I||\mathbf{n}_j\cdot\bm{\omega}_I|}{||\mathbf{p}_j-\mathbf{p}_i||^2_2}}_{\text{rough estimate of decay}}, 
\end{equation}
where $\text{clamp}(\cdot)$ denotes clamping the values at three channels (RGB) into a single scalar value. In the forward pass, we use the converted illumination from the RGB color space, while in the gradient calculation, we use the norm of gradients for all three channels.

However, we need to estimate the next event for all $N$ kernels simultaneously. It is of $\mathcal{O}(N^2)$ complexity and explicitly constructing the probability matrix is costly. In fact, we could sample $j$ based on $p_j$ without seeing all the probabilities simultaneously and without knowing the normalization factor in advance. Specifically, for the $i^\text{th}$ kernel, we stream it with the calculated unnormalized $p_j$ for every other kernel only once and sample $j\sim p_j$. This is known as the online streaming weighted sampling and well solved by the \emph{weighted reservoir sampling} \cite{wrs}. 

We further improve the computational complexity of the next event estimation by taking inspiration from light cuts \cite{light-cut}. The choice of the $i$ (Line~\ref{alg:select-start}-\ref{alg:select-end} in Alg.~\ref{monte-carlo-solver}) can also be flexible to improve the efficiency. Please find more details in Appendix~\ref{app:mc}.

\paragraph{Hybrid Solver.} As mentioned, at each step of progressive refinement, the remaining sub-problem of solving outgoing radiance given unshot radiances is in itself again a complete light transport problem. Therefore, we could first apply the progressive refinement to estimate the outgoing radiance from all light sources to all other kernels, which is conceptually similar to the direct light factorization and light tracing%
, and then apply our Monte-Carlo solver to estimate the indirect illumination. This approach is much more efficient and has lower variance for a similar compute time, compared to simply using the Monte-Carlo solver to estimate the global illumination. %

\begin{figure}
    \centering
    \includegraphics[width=\linewidth]{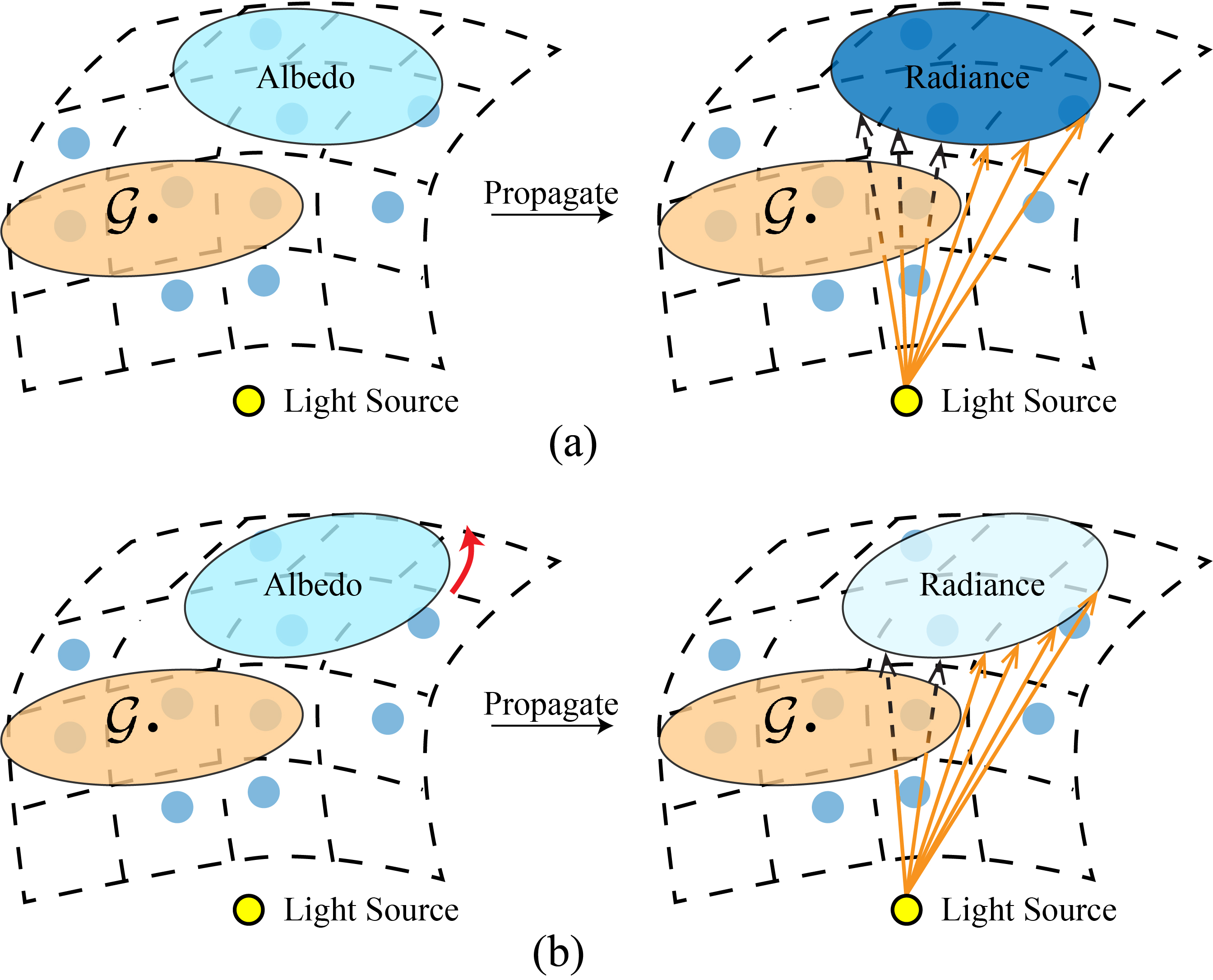}
    \caption{Illustration of the sensitivity of outgoing radiance to small geometric changes. \textbf{(a)} Due to the {constant outgoing radiance assumption}%
    , the propagated outgoing radiance for the blue kernel is the averaged response to the integral of propagated incoming radiance along all possible directions from the light source. Some rays are occluded, while others are not. \textbf{(b)} After a small geometric change (rotate and translate a little bit), the visibility for different rays could change substantially, which results in an inescapable change to the outgoing radiance.}
    \label{fig:occlusion-var}
    \vspace{-8pt}
\end{figure}

\section{Approximations}
\label{method-approx}
After establishing the whole framework and the solvers, as illustrated in Fig.~\ref{fig:occlusion-var}, we find that, due to our {first assumption that the material, emission and outgoing radiance are constant within the support}%
, the final outgoing radiance of a kernel is the average of all possible rays propagating the radiance for that kernel. These rays could have a large variance in terms of visibility that is quite sensitive to geometric changes. In turn, the final outgoing radiance of a kernel is sensitive to small geometric disturbance. This may make the optimization of both geometry and material much harder since we could optimize the geometry from a set of randomly initialized points, which implies potential strong geometric noise during the optimization. Besides, the optimization of the geometry is also coupled with that of the material.
Therefore, we are motivated to look for approximations that can relax the dependency between geometry and material while being much more efficient.

Inspired by \citet{radiosity_hierarchical_glossy}, we propose a variance-free and efficient ``center-to-center'' approximation, which we only consider the ray connecting the central points of the Gaussian surfels. In Eqn.~\eqref{eqn:before-final-radiosity}, we assume that the BRDF term, outgoing radiance term and the decay term are almost constant inside the integral: %

\begin{equation}
\label{eqn:radiosity-approx-1}
\begin{aligned}
    \mathbf{B}^c_i \approx& \mathbf{E}^c_i + \alpha_i(\mathbf{p}_i)\frac{\int_{P_i}\kappa_id\mathbf{x}}{\Lambda_i}\sum_{j=1}^{N}{\mathbf{f}_i(\bm{\omega}_I)} {(\mathbf{Y}(\bm{\omega}_I)^T\mathbf{B}^c_j)} (\int_{P_j}\alpha_j(\mathbf{x})d\mathbf{x}) \\
    &(\prod_{k=1}^{j-1}(1-\alpha_k)\frac{|\mathbf{n}_i\cdot\bm{\omega}_I||\mathbf{n}_j\cdot\bm{\omega}_I|}{||\mathbf{p}_i-\mathbf{p}_j||_2^2}), 
\end{aligned}
\end{equation}
where $\bm{\omega}_I=\frac{\mathbf{p}_i-\mathbf{p}_j}{||\mathbf{p}_i-\mathbf{p}_j||_2^2}$, $\frac{\int_{P_i}\kappa_id\mathbf{x}}{\Lambda_i}=1$, and we also have the closed-form solution for $\int_{P_j}\alpha_jd\mathbf{x}$, which is detailed in Appendix~\ref{app:appr}.

To compensate for the error in the approximation, we further assign an optimizable scaling factor $\lambda_i\in\mathbb{R}_{+}$ to each $i^\text{th}$ kernel.
We now absorb the $\alpha_i$ term into the $\mathbf{f}_i(\bm{\omega}_I)$ term, and Eqn.~\eqref{eqn:radiosity-approx-1} can be simplified as:
\titledeq{\xspace forward rendering equation (approximated)}{primalForeground}{primalBackground}{
\begin{equation}
\begin{aligned}
    &\mathbf{B}^c_i = \mathbf{E}^c_i + \sum_{j=1}^{N}{\mathbf{f}_i(\bm{\omega}_I)} {(\mathbf{Y}(\bm{\omega}_I)^T\mathbf{B}^c_j)} V_{ji}, \\ 
    &\text{where }V_{ji}=\lambda_j(\int_{P_j}\alpha_jd\mathbf{x}) (\prod_{k=1}^{j-1}(1-\alpha_k)\frac{|\mathbf{n}_i\cdot\bm{\omega}_I||\mathbf{n}_j\cdot\bm{\omega}_I|}{||\mathbf{p}_i-\mathbf{p}_j||_2^2})
\end{aligned}
\end{equation}
}
In the equation, the fused decay term $V_{ji}$ is updated to absorb $\lambda_j\int_j\alpha_jd\mathbf{x}$ terms, so that the gradient calculation and the solvers can be easily updated by using the redefined symbols and removing the double integrals without altering the general framework. %

\begin{figure*}
    \centering
    \includegraphics[width=\linewidth]{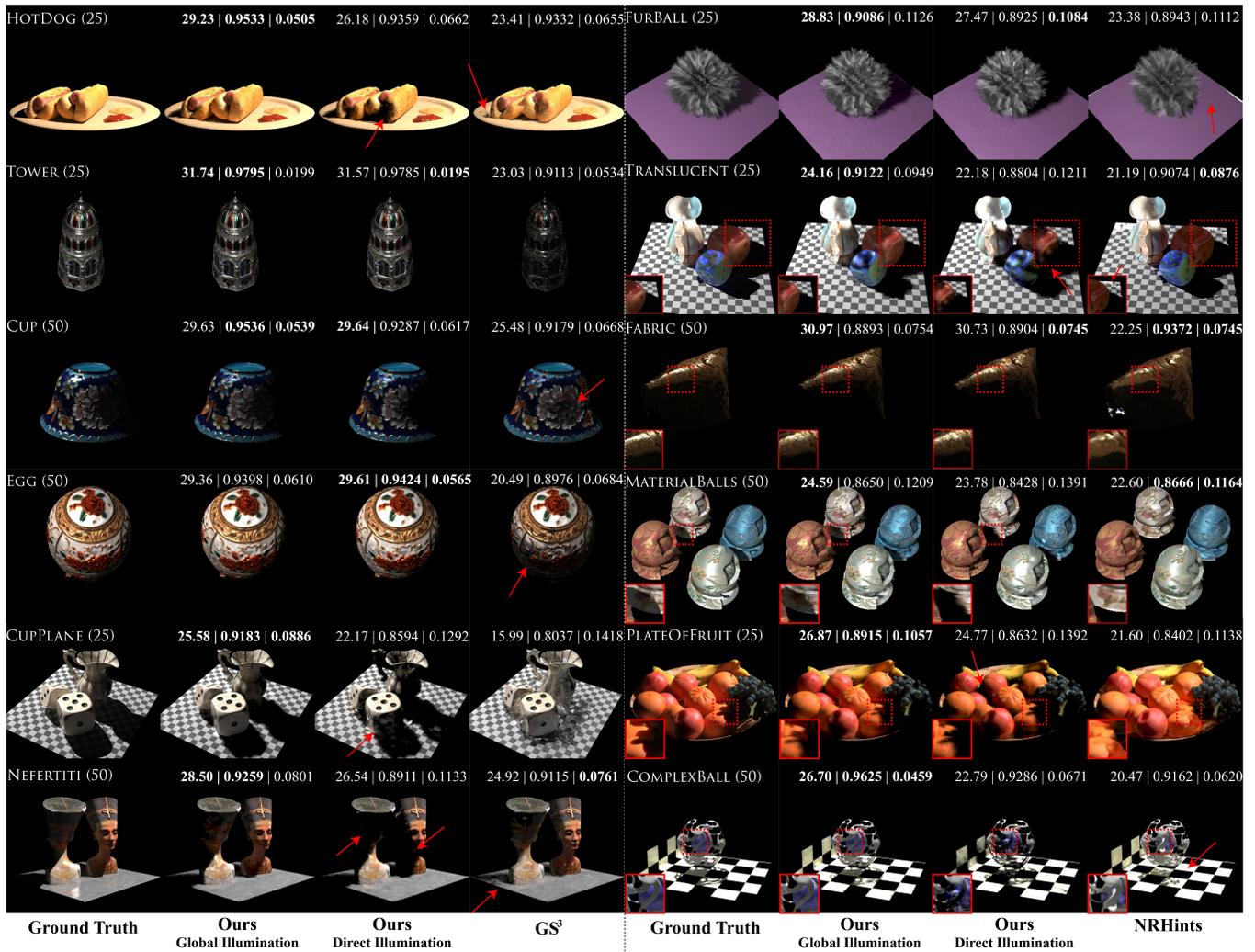}
    \caption{Qualitative comparison of relighting quality on the $12$ scenes from ``RenderCapture'' and ``Synthetic'' datasets \cite{gs3}{ and our self-made scenes}. We label the name of the scene and the number of images used for training at the left-top corner of the ground-truth image, and the PSNR, SSIM, LPIPS metrics, \emph{resp.}, on the top for measuring relighting quality of rendered images from different methods. We bold the best metrics and non-obvious differences are highlighted by red arrows and insets. Notice that SSIM and LPIPS metrics are less sensitive to shadow errors.
    The \textsc{Nefertiti} scene is adapted from \textsc{Nefertiti's bust} by C. Yamahata\textcopyright Sketchfab.
    }
    \label{fig:view-synthesis}
\end{figure*}

\section{Results}

\paragraph{Additional Implementation Details.} We set the emissions of all kernels except for special kernels representing the light sources to be \emph{zero} to avoid causing ambiguity with the light sources. When the light source is unknown, we use an optimizable environment map with resolution $16\times32$ approximated by distant point light sources without inverse square falloff. %
The hyperparameters are set almost the same as \citet{gfsgs}'s work. Besides, we observe the original densification strategy in \cite{3DGS} could cause significant disruption in the geometry and then view synthesis, which makes the optimization diverge. Therefore, we adopt \citet{evsplit}'s strategy to replace the original splitting strategy to preserve the geometry as much as possible and take inspiration from \citet{nd_Gaussian} to enforce clone-only densification after half of the densification procedure. {During inference, we double the time step in the hybrid solver to solve the light transport.}

The framework is implemented using PyTorch \cite{pytorch} and OptiX \cite{optix}. We conduct our experiments on an NVIDIA 6000 Ada GPU. The optimization takes $5-10$ hours to take global illumination into account. In comparison, the optimization takes about $20$ minutes if only direct illumination is considered. 

\paragraph{Datasets.} We evaluate the sparse-view relighting, view synthesis, and geometry reconstruction capabilities of our method using ground-truth high dynamic range images for supervision. Specifically, for relighting, we use the ``RenderCapture'' and ``Synthetic'' datasets of \citet{gs3} {and four more of our self-made scenes featuring rich indirect illumination}. Our method currently does not support differentiating through the camera parameters, and we therefore do not use datasets containing camera errors. We modify the datasets{ of \citet{gs3}} such that only a small number of training images are retained ($25,50$ images), while the other $\sim2000$ images are used for testing. Each image is associated with varying known lighting conditions.
This is used to demonstrate the benefits of having physically-based rendering compared to data-driven methods. In total, there are $24$ scenes with known point light sources. For view synthesis and geometry reconstruction, we use the ``Stanford-ORB'' \cite{stanfordorb} dataset, which contains $42$ scenes with unknown environment map as the light source.
Notably, these datasets do not provide a set of points for initialization, and we therefore use randomly initialized points distributed within a cube for all baselines.

\begin{figure*}
    \centering
    \includegraphics[width=\linewidth]{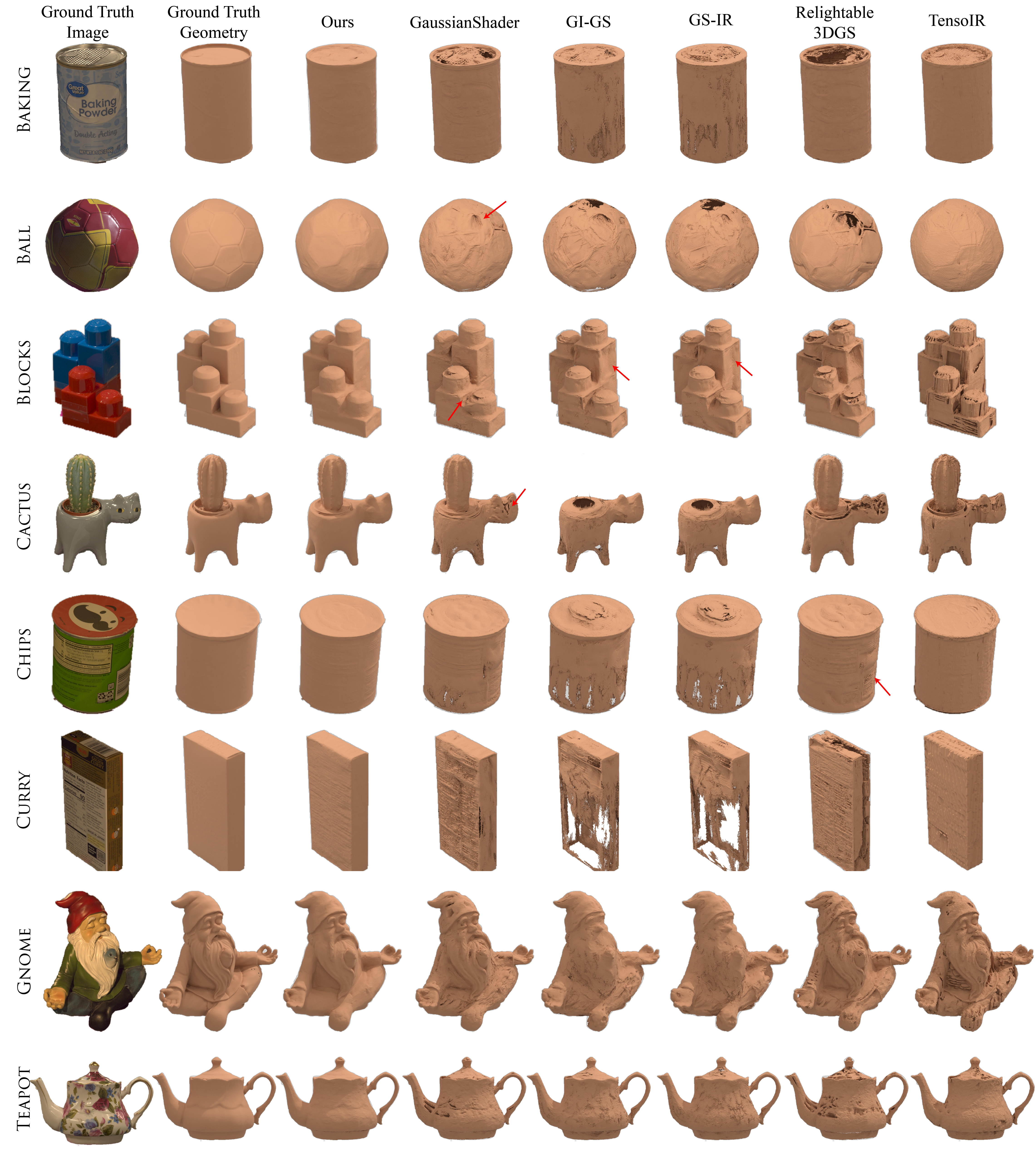}
    \caption{Qualitative comparison of geometry reconstruction on the Stanford-ORB dataset \cite{stanfordorb}. The name of the object is labeled at the left. Non-obvious differences are highlighted by red arrows.}
    \label{fig:stanford-orb}
\end{figure*}

\paragraph{Baselines.} For relighting{ alone}, we compare with NRHints \cite{nrhints} and GS$^3$ \cite{gs3}, which are representative of NeRF-based and Gaussian-splatting-based methods, respectively. For {all} view synthesis, geometry reconstruction{ and relighting}, we compare with methods tailored for objects lit by environment maps, including TensoIR \cite{tensoir}, GS-IR \cite{GSIR}, GI-GS \cite{GIGS}, Relightable3DGS \cite{relightable3DGS}, and GaussianShader \cite{Gaussianshader}. 
Notably, we do not compare with Gaussian splatting methods which support ray tracing as they still rely on the radiance fields assumption, and it is non-trivial to extend them into full differentiable light transport. We also do not compare with Mitsuba3 \cite{Mitsuba3} as it cannot easily solve the highly under-constrained inverse problems from scratch which are exactly our cases.

We separately list ours but with only direct illumination effects during training and inference as our baseline.
We adopt the TSDF fusion \cite{tsdf, open3d} to extract the mesh for all Gaussian-based methods with the same hyperparameters for fairness.
We mask out the background and apply the masking loss for compatible baselines for a fair comparison. For some baselines requiring a white background for training, we follow this convention while switching the background into a black one for consistent demonstration across the paper.

\begin{table}[t]
\caption{Quantitative evaluation of the synthesized relit test image quality on the ``RenderCapture'' and ``Synthetic'' datasets{ and our self-made scenes}, given $25$ and $50$ training images. The best metrics are emphasized in bold and the second best metrics are emphasized by an underline.}
\setlength{\tabcolsep}{2pt}
\small
\centering
\begin{tabular}{l|ccc|ccc}
\hline
 & \multicolumn{3}{c|}{\textbf{25-view}} & \multicolumn{3}{c}{\textbf{50-view}} \\
\multicolumn{1}{l|}{\textbf{Method}} & {PSNR$\uparrow$} & {SSIM$\uparrow$} & {LPIPS$\downarrow$} & {PSNR$\uparrow$} & {SSIM$\uparrow$} & {LPIPS$\downarrow$} \\ \hline
{GS$^3$\cite{gs3}} & {21.80} & {0.8607} & {0.1175} & {24.67} & {0.8992} & 0.0888 \\
NRHints \cite{nrhints} & 23.18 & \textbf{0.9050} & \textbf{0.0871} & 25.41 & \textbf{0.9273} & \textbf{0.0741} \\
\hline
Ours \emph{(Direct Illumination)} & \underline{24.36} & 0.8794 & 0.1085 & \underline{26.24} & 0.8982 & 0.0942 \\
Ours \emph{(Global Illumination)} & \textbf{25.29} & \underline{0.8950} & \underline{0.0985} & \textbf{27.43} & \underline{0.9151} & \underline{0.0830} \\
\hline
\end{tabular}
\label{table:view-synthesis}
\end{table}

\paragraph{Metrics.} We use the PSNR, SSIM and LPIPS \cite{lpips} error metrics for evaluating the view synthesis quality compared against the ground-truth images. We also use the Chamfer distance ($\times10^3$), denoted as ``CD'', to measure the geometry reconstruction quality compared to the ground-truth meshes. 

\subsection{Qualitative Comparison}
\label{exp:qualitative}
In terms of relighting, as shown in Fig.~\ref{fig:view-synthesis}, ours with global illumination mostly achieves the best relighting quality, as evidenced by the PSNR metric. We select test views where indirect illumination is obvious if present. Even for the challenging case ``Translucent'', ours still achieves reasonable results despite a somewhat less translucent appearance than the reference{ due to the simplified material model}. For the ``HotDog'', ``Cup'' {and ``CupPlane''} cases, GS$^3$ misses the shadows indicated by the red arrows. For the ``Tower'', ``Egg'' {and ``Nefertiti''} cases, GS$^3$ produces visible artifacts such as wrongly darkened surfaces{ or incomplete shadows}. 
For the ``FurBall'', ``MaterialBall''{, ``PlateOfFruit'' and ``ComplexBall''} cases, NRHints also misses the shadows. For ``Translucent'', ``Fabric''{ and ``ComplexBall''} cases, NRHints produces visible texture errors indicated by the insets. 
However, SSIM and LPIPS metrics are less sensitive to shadow errors and even prefer smooth results occasionally. For the``MaterialBall'' scene, NRHints has the best SSIM and LPIPS metrics despite visible shadow errors. For `the `Fabric'' scene, NRHints synthesizes too smooth texture without details, resulting in a low PSNR metric but surprisingly high SSIM and low LPIPS metrics.
Ours with direct illumination cannot synthesize indirect illumination effects while being on par with ours with global illumination when the indirect illumination is negligible, which will be further elaborated later in the ablation study.

In terms of geometry reconstruction, as shown in Fig.~\ref{fig:stanford-orb}, other Gaussian-splatting-based methods produce meshes with obvious holes or meshes which overfit to the textures, such as words at the back of the ``Curry'' scene. TensoIR sometimes fails to produce smooth meshes while exhibiting visible holes or blocky artifacts, as evidenced in ``Blocks'', ``Cactus'', ``Gnome'' and ``Teapot'' scenes. In contrast, ours produces detailed meshes without cracks or holes.

\begin{table}[t]
\caption{Quantitative evaluation of the novel view synthesis quality and geometry reconstruction quality on the Stanford-ORB dataset. The chamfer distance is denoted as ``CD'' ($\times10^3$). The best metrics are emphasized in bold and the second best metrics are emphasized by an underline.}
\small
\centering
\begin{tabular}{l|ccc|c}
\hline
                & \multicolumn{3}{c|}{\textbf{NVS}}              & \multicolumn{1}{c}{\textbf{Geo}}             \\
\textbf{Method} & {PSNR$\uparrow$} & {SSIM$\uparrow$} & {LPIPS$\downarrow$} & {CD$\downarrow$} \\ \hline
TensoIR \cite{tensoir} & 33.03 & \underline{0.9839} & 0.0314 & 0.503 \\
GS-IR \cite{GSIR} & 35.59 & 0.9784 & 0.0398 & 0.308 \\
GI-GS \cite{GIGS} & 34.78 & 0.9752 & 0.0412 & 0.306 \\
Relightable3DGS \cite{relightable3DGS} & 28.87 & 0.9506 & 0.0329 & 0.280 \\
GaussianShader \cite{Gaussianshader} & \underline{37.44} & \textbf{0.9894} & \textbf{0.0210} & \underline{0.224} \\  \hline
Ours & \textbf{39.30} & \textbf{0.9894} & \underline{0.0251} & \textbf{0.173} \\ \hline
\end{tabular}
\label{table:geometry-reconstruction}
\end{table}

\begin{table}[t]
\caption{{Quantitative evaluation of the relighting quality on the Stanford-ORB dataset. The best metrics are emphasized in bold and the second best metrics are emphasized by an underline.}}
\small
\centering
\begin{tabular}{l|ccc}
\hline
 & \multicolumn{3}{c}{\textbf{Relighting}}   \\
\textbf{Method} & {PSNR$\uparrow$} & {SSIM$\uparrow$} & {LPIPS$\downarrow$} \\ \hline
GS-IR \cite{GSIR} & \underline{31.16} & 0.9341 & 0.0548 \\
GI-GS \cite{GIGS} & 28.73 & 0.9421 & 0.0534 \\
Relightable3DGS \cite{relightable3DGS} & 28.68 & 0.9377 & 0.0602 \\
GaussianShader \cite{Gaussianshader} & 30.57 & \underline{0.9627} & \underline{0.0392} \\  \hline
Ours & \textbf{32.72} & \textbf{0.9721} & \textbf{0.0365} \\ \hline
\end{tabular}
\label{table:relighting}
\end{table}

\subsection{Quantitative Comparison}
In terms of relighting, as shown in Table~\ref{table:view-synthesis}, ours with global illumination achieves the best view synthesis quality under unseen lighting conditions and viewpoints indicated by the PSNR metrics. In terms of SSIM and LPIPS metrics, ours are slightly worse than NRHints. We argue that it is because NRHints tends to synthesize smooth results due to the underlying continuous representation of radiance fields, i.e., MLPs, and SSIM and LPIPS are less sensitive to shadow errors, as explained in Sec.~\ref{exp:qualitative}. Ours with direct illumination misses the indirect illumination effects, leading to worse metrics and significantly worse visual effects when the indirect illumination is obvious.

In terms of %
novel view synthesis,  geometry reconstruction {and relighting on the Stanford-ORB dataset}, as shown in Tables~\ref{table:geometry-reconstruction}, \ref{table:relighting}, ours achieves the best PSNR and chamfer distance metrics across baselines, demonstrating the effectiveness of the proposed pipeline. On SSIM and LPIPS metrics {of novel view synthesis}, ours are {on par with or} slightly worse than GaussianShader. However, as discussed before, SSIM and LPIPS metrics are less sensitive to shading errors.

\begin{figure}[t]
    \centering
    \includegraphics[width=\linewidth]{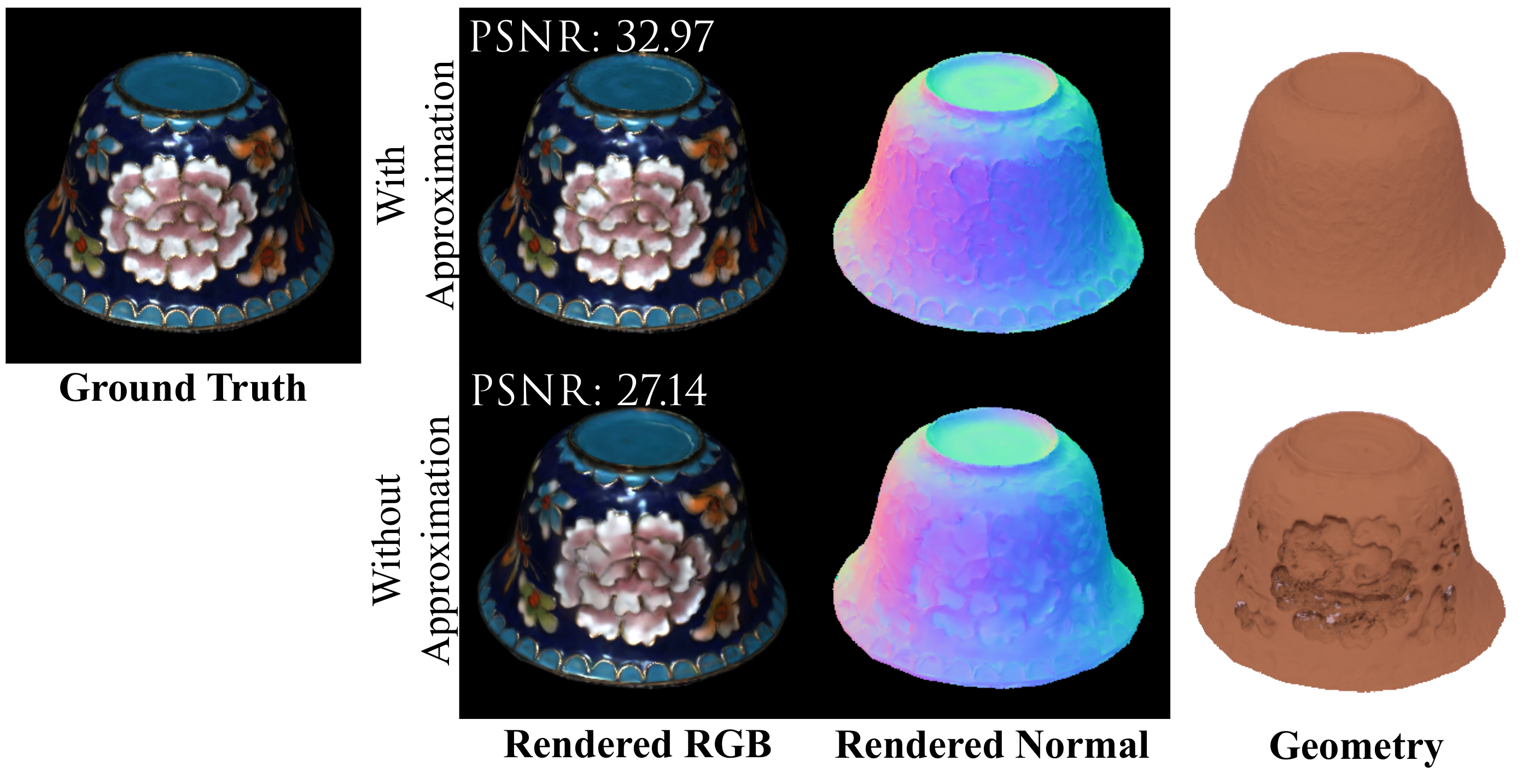}
    \caption{Ablation study on the effects of the proposed approximation. Without the approximation in the second row, the optimization is much harder, leading to worse view synthesis quality (the flower is wrongly darker) evidenced by the PSNR metric and noisy geometry reconstruction, compared to ours with the approximation in the first row.}
    \label{fig:ablation-apprx}
\end{figure}

\begin{figure}[t]
    \centering
    \includegraphics[width=\linewidth]{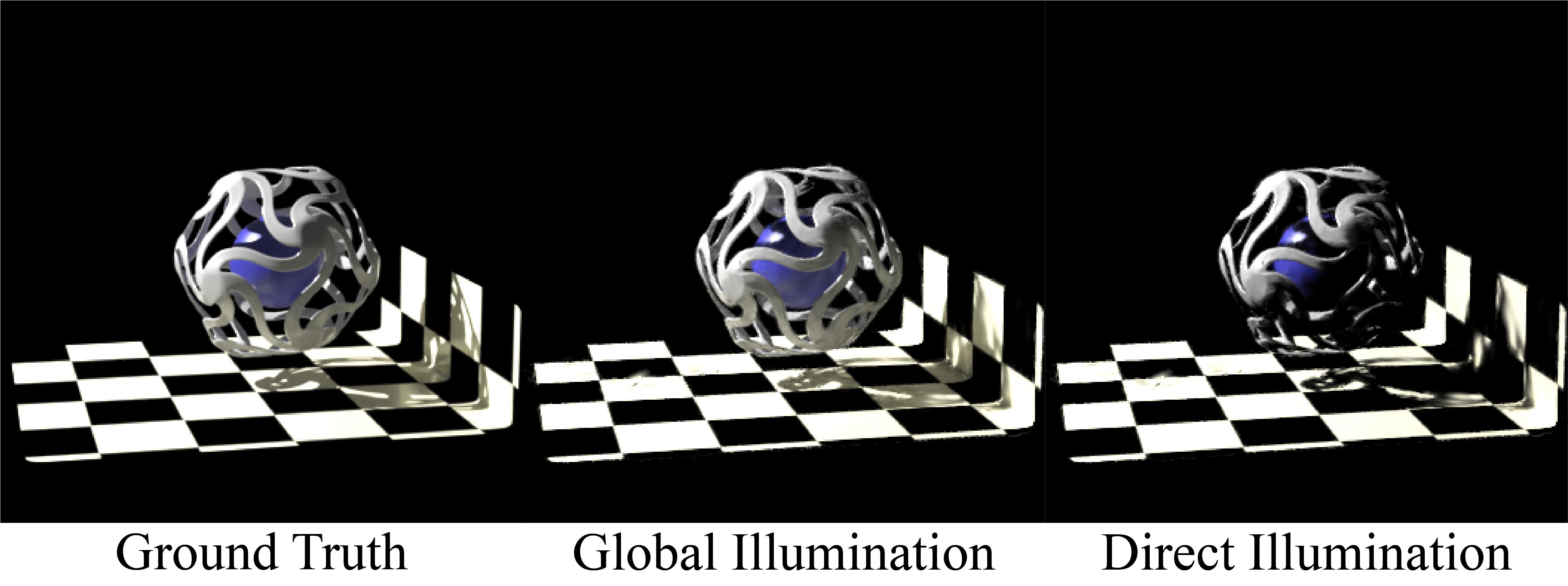}
    \caption{Another comparison between the direct illumination and global illumination using the ``ComplexBall'' scene from \cite{nrhints} {with dense training images}. Notably, there are {still} significant visual differences between the global illumination result and the direct illumination result.}
    \label{fig:direct_versus_global}
\end{figure}

\subsection{Ablation Study}
\label{sec-exp-ablation}
\paragraph{Approximation.}
We conduct an ablation study about our proposed approximation in Sec.~\ref{method-approx}. Specifically, we use the ``Cup'' scene (50 training images setup) from the RenderCapture dataset \cite{gs3}. We remove the approximations, and consequently the learnable compensation factors as well, for both training and testing, as our baseline. We use Monte-Carlo estimation to estimate the double integral and importance sample the $\mathbf{x}$ and $\mathbf{x'}$ using the underlying Gaussian distributions in Eqn.~\ref{eqn:final-radiosity}.

As explained in Sec.~\ref{method-approx}, the optimization of both geometry and material is a tricky task, and their mutual strong dependency makes it even worse. As shown in Fig.~\ref{fig:ablation-apprx}, due to the strong dependency and noise in both forward rendering and gradient calculation without the approximation, the optimization is harder, leading to poorer view synthesis quality and geometry reconstruction. In the test set of $2000$ images, the baseline without approximation achieves $26.44$, $0.9255$, $0.0681$ for PSNR, SSIM, LPIPS, respectively. In contrast, ours achieves $30.01$, $0.9542$, $0.0430$ for PSNR, SSIM, LPIPS, demonstrating a significant performance boost by relaxing the dependency between the optimization of geometry and material.

\paragraph{Global illumination versus direct illumination.} We conduct comparison between global illumination effects and direct illumination effects both qualitatively and quantitatively, as shown in Fig.~\ref{fig:view-synthesis} and Table~\ref{table:view-synthesis}. The indirect illumination effects are relatively subtle in the object-centric cases but they require efforts to accurately reproduce. Qualitatively, in the ``HotDog'', ``Translucent'', ``MaterialBall'', {``CupPlane'', ``Nefertiti'', ``PlateOfFruit'' and ``ComplexBall''} scenes, the direct illumination misses the reflected rays, leading to incorrectly unlit regions emphasized by red arrows or insets, and therefore compromised relighting quality. We further show {another} example in Fig.~\ref{fig:direct_versus_global} using the ``ComplexBall'' scene %
from \cite{nrhints}{ even with dense training images (2000 images)}.
Quantitatively, ours with direct illumination also performs worse than ours with global illumination, but the margin is not large because the indirect illumination effects are local and subtle.

\begin{figure}[t]
    \centering
    \includegraphics[width=\linewidth]{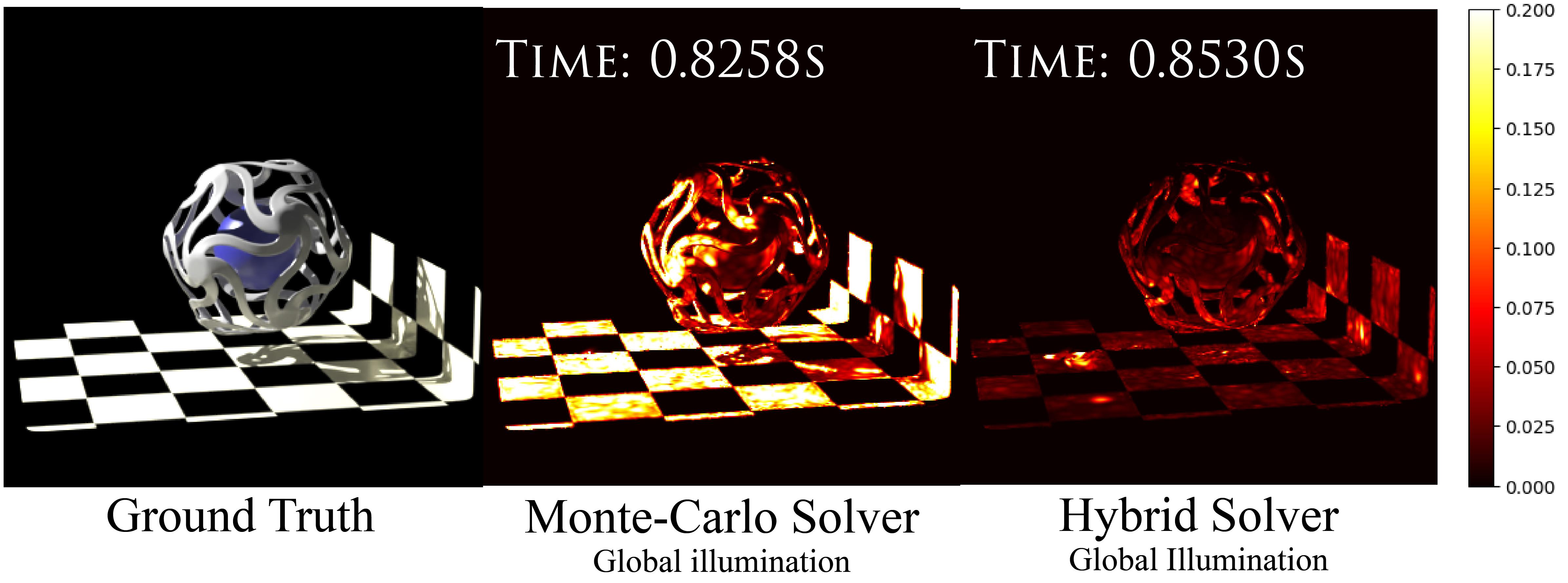}
    \caption{Comparison between the hybrid solver and Monte-carlo solver for the ``ComplexBall'' scene with $\sim89$K primitives. The variances of rendered images with global illumination using different solvers are demonstrated by colored heatmaps (colorbar is visualized on the right). The time cost for calculating the light transport is labeled at the left-top corner.}
    \label{fig:variance}
\end{figure}

\begin{table}[t]
\caption{Ablation study for batch size in the optimization. We use the novel view synthesis and geometry reconstruction tasks on the Stanford-ORB dataset for evaluation. The best metrics are emphasized in bold.}
\small
\centering
\begin{tabular}{l|ccc|c}
\hline
                & \multicolumn{3}{c|}{\textbf{NVS}}              & \multicolumn{1}{c}{\textbf{Geo}}             \\
\textbf{Method} & {PSNR$\uparrow$} & {SSIM$\uparrow$} & {LPIPS$\downarrow$} & {CD$\downarrow$} \\ \hline
Ours (Batch size $4$) & \textbf{40.05} & \textbf{0.9908} & \textbf{0.0218} & 0.190 \\ 
Ours (Batch size $1$) & 39.30 & 0.9894 & 0.0251 & \textbf{0.173} \\ 
\hline
\end{tabular}
\label{table:ablation-batch}
\end{table}

\paragraph{Hybrid solver versus Monte-Carlo solver.} As discussed before, the hybrid solver using progressive refinement and Monte-Carlo solver could significantly reduce the variance with slightly more time cost compared to the Monte-Carlo solver only. We set up an experiment for comparison using the ``ComplexBall'' scene with a known point light source as shown in Fig.~\ref{fig:variance}.

We use the Monte-Carlo solver and hybrid solver for solving the light transport respectively with the same time step $T=64$. We run each solver for $20$ times and calculate the variance of these $20$ rendered images, which is visualized as a heatmap. 
Even in the case of having only one point light source, factoring out the direct illumination and then solving the light transport for indirect illumination using the hybrid solver performs much better than solving the light transport for full global illumination (including direct lighting) using the Monte-Carlo solver.
The hybrid solver significantly reduces the variance with slightly more time cost, compared to using the Monte-Carlo solver.
Notably, if we simply use the progressive refinement to solve the whole light transport, we need to theoretically spend at least $0.0183\text{s}\times 100\text{K}\approx0.5$ hour to finally converge.

We further train the scene with the pure Monte-Carlo solver. Using the Monte-Carlo solver instead of the hybrid solver for training decreases the PSNR by $0.25$ db on test views, while increasing the training time by $35.77$\%. The higher variance of the Monte-Carlo solver introduces more kernels during densification, while achieving lower synthesis quality.

\begin{figure}[t]
    \centering
    \includegraphics[width=\linewidth]{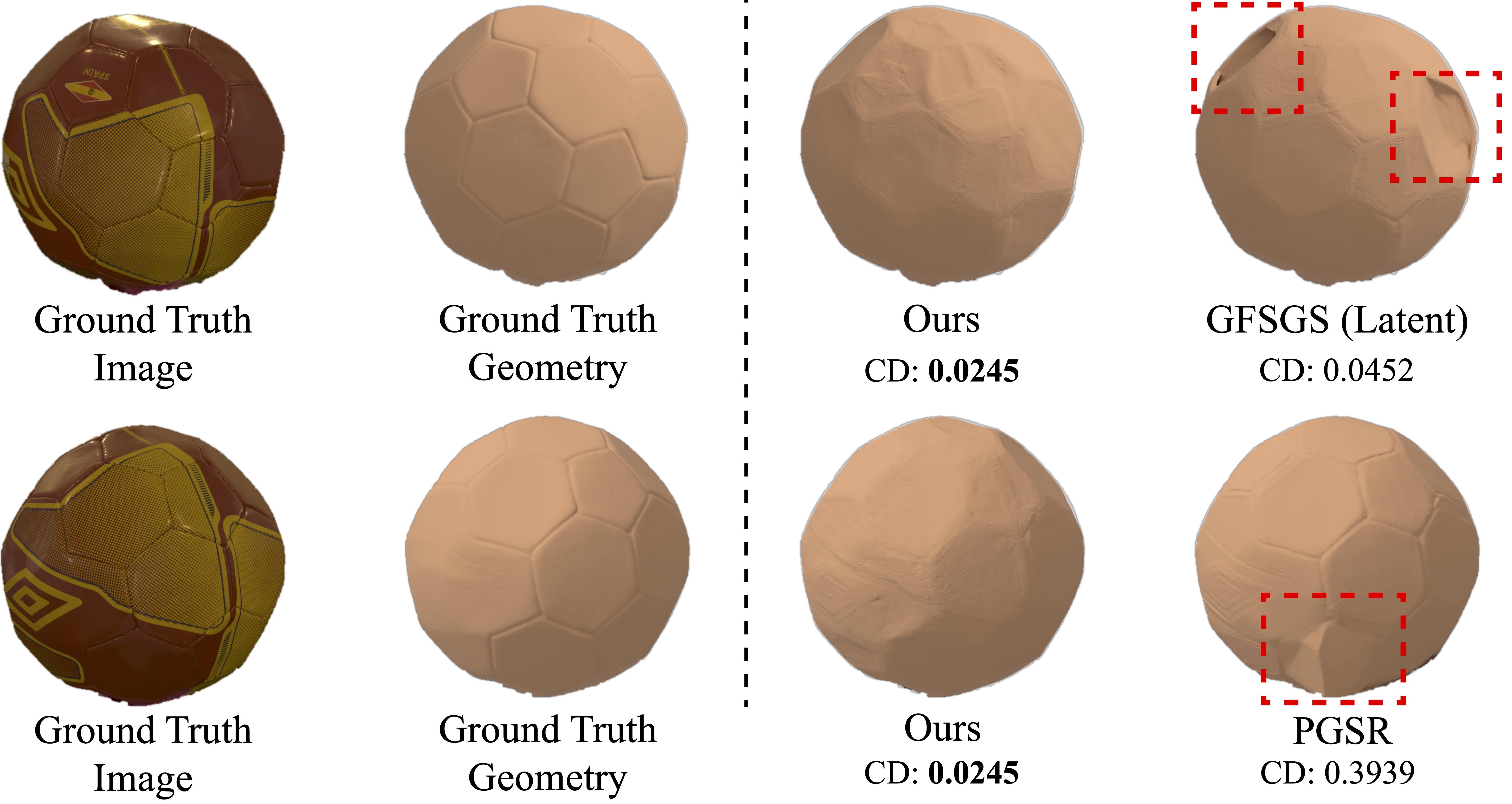}
    \caption{Ablation study for geometry disambiguation. We compare ours with geometry reconstruction methods using the radiance fields, i.e., GFSGS \cite{gfsgs} with latent representation for color and PGSR \cite{pgsr}. Wrong geometry is emphasized by the red rectangles. We also label the chamfer distance ($\times10^3$), denoted as ``CD'', under each method and bold the best metric.}
    \label{fig:geometry-disambiguation}
\end{figure}

\paragraph{Proposed gradient calculation versus auto-differentiation.} We compare the efficiency between the proposed gradient calculation and the auto-differentiation using PyTorch \cite{pytorch}. Specifically, we still use the ``ComplexBall'' scene in Fig.~\ref{fig:variance} with $\sim89$K primitives. We use time step $T=32$ instead of default $64$, as otherwise the auto-differentiation would run out of memory. With our proposed gradient calculation, our full backward pass takes $0.26$s. In contrast, auto-differentiation takes $2.34$s, and has a $28.25$ GB larger memory footprint owing to the need for storing operations in the forward pass. Therefore, our proposed gradient calculation effectively reduces the time consumption to one tenth, and eliminates the need for recording forward-pass operations, to significantly reduce the memory footprint. We have also verified the correctness of our gradient calculation by comparing the gradients given by our method and those given by auto-differentiation. %

We also experiment with removing the proposed gradient calculation for the geometry, i.e., central points, scalings, rotations, etc., in our framework and only use the gradients from the differentiable tile rasterizer as it might be the case that the gradients from the differentiable tile rasterizer dominate. However, in that case, the optimization diverges.

\paragraph{Use a batch of training images at each optimization step.} We conduct quantitative experiments on the Stanford-ORB dataset \cite{stanfordorb} to study the batch optimization enabled by the \emph{view-independent} rendering. Specifically, during training, we could only propagate the radiance once to calculate the light transport and render multiple training images for supervision to greatly speed up the training process. We compare ours trained with batch size $4$ with ours trained with batch size $1$. The learning rates are also scaled following \cite{zhao2024scaling}. As shown in Table~\ref{table:ablation-batch}, ours with batch size $4$ only spends a bit more than one fourth of the original training time due to overhead {and the view synthesis quality improves} but the geometry reconstruction quality {is} slightly compromised. However, ours with batch size $4$ is still better than the baselines shown in Table~\ref{table:geometry-reconstruction}.

\begin{table}[t]
\caption{Ablation study for the geometry disambiguation using the quantitative evaluation of the geometry reconstruction quality on the Stanford-ORB dataset. GFSGS \cite{gfsgs} using the spherical harmonics representation for color is labeled as ``GFSGS (SH)'', while GFSGS with the latent representation for color is labeled as ``GFSGS (Latent)''. The best metric is emphasized in bold.}
\small
\centering
\begin{tabular}{c|cccc}
\hline
\textbf{Method} & GFSGS (SH) & GFSGS (Latent) & PGSR & Ours \\
\hline
CD $\downarrow$ & 0.361 & 0.273 & 0.314 & \textbf{0.173} \\
\hline
\end{tabular}
\label{table:ablation-geometry-reconstruction}
\end{table}

\paragraph{Geometry disambiguation.} We illustrate the effectiveness of enabling the light transport for disambiguating the geometry reconstruction. Specifically, we compare with GFSGS \cite{gfsgs} and PGSR \cite{pgsr} as representative works for geometry reconstruction from a set of calibrated images using radiance fields. Even though \citet{dipole}'s work is also a state-of-the-art geometry reconstruction method, it requires special initialization that is not available by default for training images under varying lighting conditions.

We conduct quantitative and qualitative experiments using the Stanford-ORB dataset \cite{stanfordorb}. As shown in Table~\ref{table:ablation-geometry-reconstruction}, ours achieves the best geometry reconstruction quality evidenced by the lowest chamfer distance metric. Notably, we even beat GFSGS with the latent representation, simply using the spherical harmonics basis in our method. As shown in Fig.~\ref{fig:geometry-disambiguation}, with the explicit light transport instead of using the assumption of radiance fields, ours avoids geometric errors, including concave or protruding blocks in PGSR and  in GFSGS with the latent representation for color. 

\begin{table}[t]
\caption{{Illustration of the relighting quality given different number of lighting conditions seen during the training (denoted as ``\# LC'') on the ``PlateOfFruit'' scene.}}
\begin{tabular}{c|ccccc}
\hline
\textbf{\# LC} & \textbf{1} & \textbf{2} & \textbf{5} & \textbf{10} & \textbf{50} \\ \hline
{PSNR}$\uparrow$                   & 22.70      & 23.74      & 27.31      & 27.69       & 27.73       \\
{SSIM}$\downarrow$                   & 0.9038     & 0.9099     & 0.9330     & 0.9341      & 0.9334      \\
{LPIPS}$\downarrow$                  & 0.0894     & 0.0835     & 0.0645     & 0.0644      & 0.0648      \\ \hline
\end{tabular}
\label{table:number_of_lighting_conditions}
\end{table}

\begin{figure*}
    \centering
    \includegraphics[width=\linewidth]{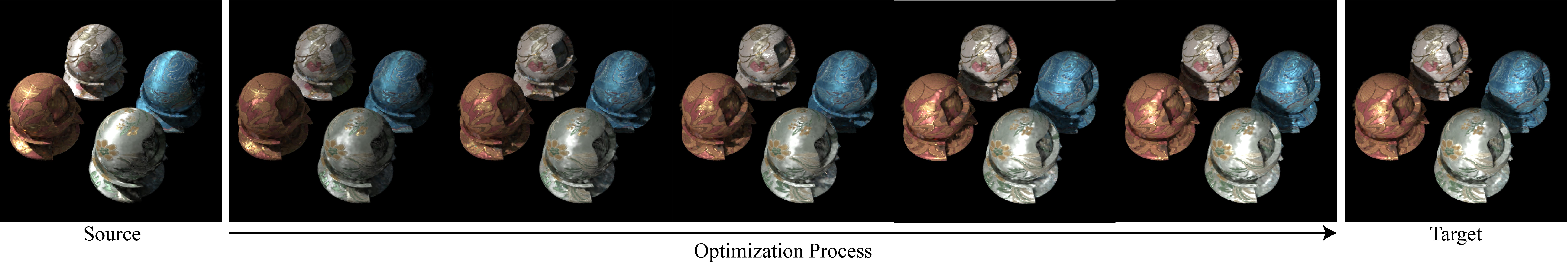}
    \caption{Illustration of optimizing the location of a point light source. Given the source image (on the far left) and the original light source location, we aim to match the target image (on the far right) by optimizing the location of a point light source based purely on the image space supervision. The movement of the light source during optimization is visualized. }
    \label{fig:light-source-optim}
\end{figure*}

\begin{figure}[t]
    \centering
    \includegraphics[width=\linewidth]{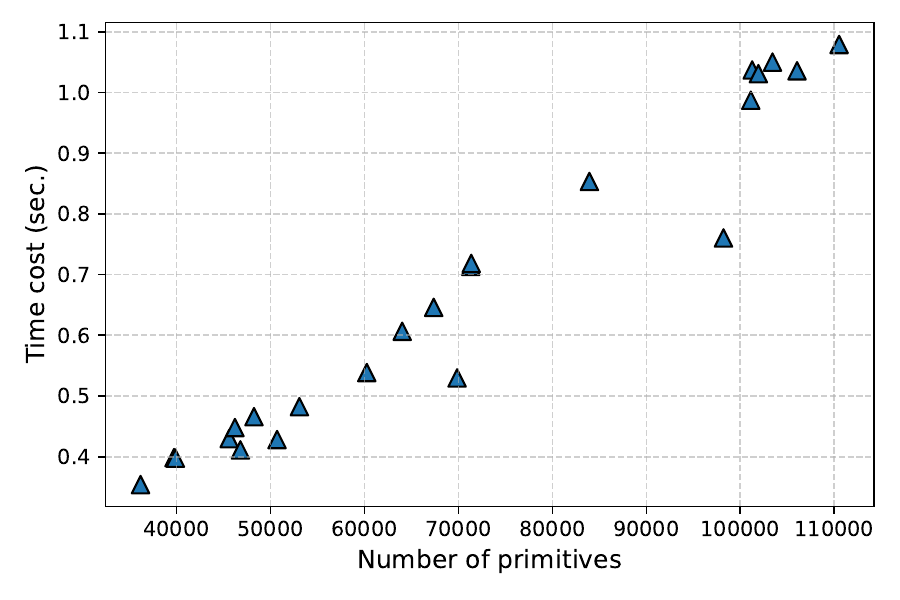}
    \caption{{Illustration of the relationship between the time cost of calculating the full light transport using the hybrid solver and the number of primitives.}}
    \label{fig:time-cost}
\end{figure}

\begin{figure}[t]
    \centering
    \includegraphics[width=\linewidth]{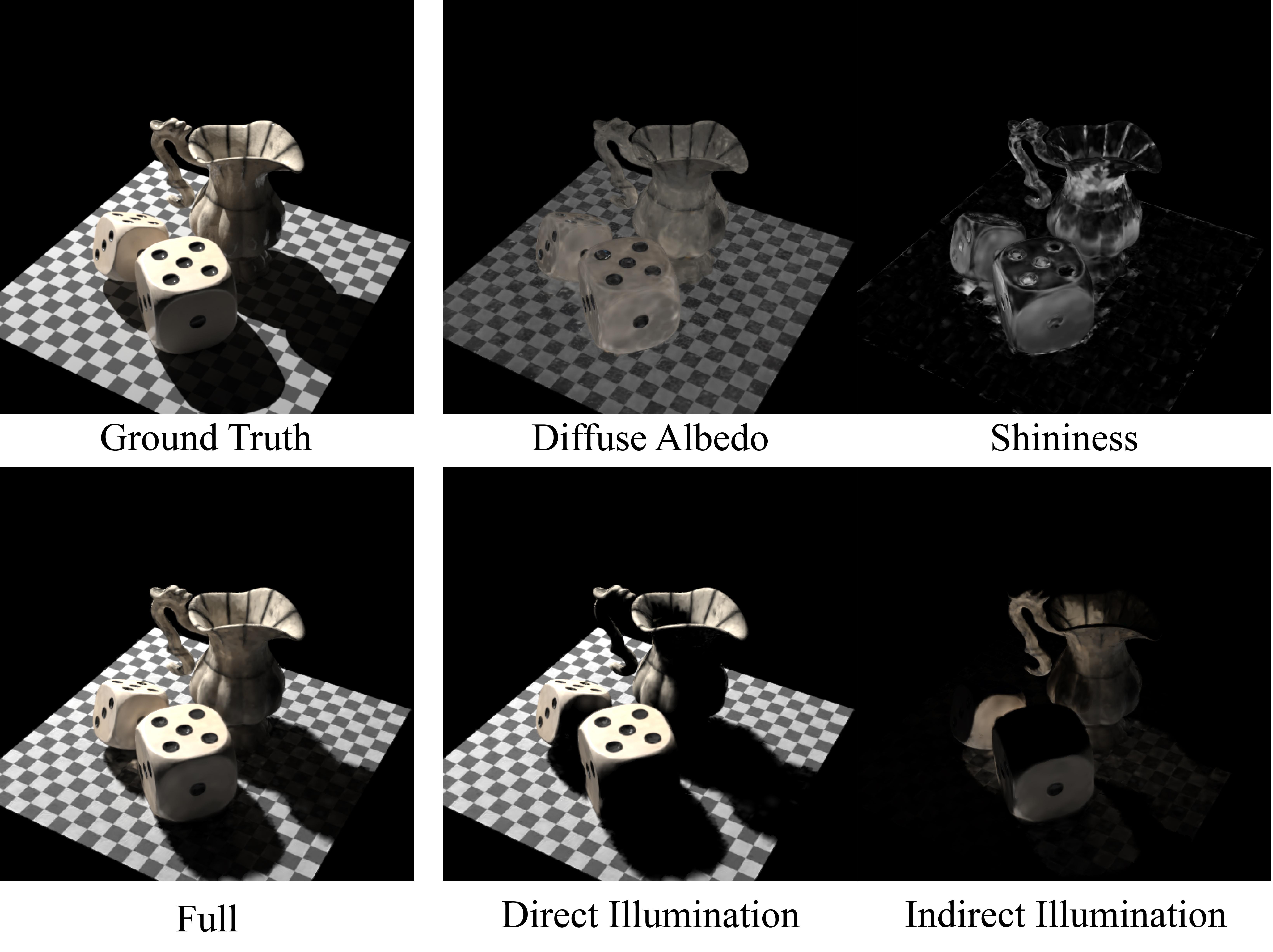}
    \caption{Illustration of decomposing a scene into different components of rendering. Given the ground truth image, the full rendered result (left-bottom one) matches with it. We are able to decompose it into diffuse albedo and shininess components (the brighter the shinier) respectively (first row), and direct illumination only and indirect illumination only (second row).}
    \label{fig:material-edit}
\end{figure}

\subsection{Additional Experiments}
Besides the main experiments we discuss above, we also conduct additional experiments to validate our method.
\paragraph{Reduce the number of different lighting conditions seen in the training.} {To be compatible with the experiment setup in GS$^3$ \cite{gs3} and NRHints \cite{nrhints}, we use varying lighting condition and varying viewpoint in each training image in the qualitative and quantitative comparisons. We further conduct an experiment using the ``PlateOfFruit'' scene with $50$ training images to study the effects of number of different lighting conditions seen in the training. Specifically, we use $1$, $2$, $5$, $10$ and $50$ different lighting conditions and the same test set with $100$ images, each of which contains a varying lighting condition. As shown in Table.~\ref{table:number_of_lighting_conditions}, perhaps surprisingly, we find that our model can effectively specify the material attributes and achieve competitive results with only $5$ different lighting conditions compared to the full setup where $50$ different lighting conditions are used.}

\paragraph{Optimize the location of light source.} When using the Stanford-ORB~\cite{stanfordorb} dataset, we optimize the environment maps which demonstrates the optimization of the emissions of light sources.
Additionally, we demonstrate optimizing the geometric property of light sources, i.e., locations, in Fig.~\ref{fig:light-source-optim} using the ``MaterialBalls'' scene from the RenderCapture dataset \cite{gs3}. Specifically, we match the target image using only L1 image space supervision by optimizing the location of the light source.
{It also demonstrates the capability of our method to continuously change the lighting condition.}

\paragraph{Time cost of calculating the light transport versus the number of primitives.} {We study the time cost of calculating the light transport using the hybrid solver given different number of primitives. We reuse the scenes in the qualitative comparison and plot the relationship in Fig.~\ref{fig:time-cost}. We identify that the time cost roughly increases linearly as the number of primitives increases. However, the time cost also depends on the specific distribution of primitives.
}

\begin{figure*}
    \centering
    \includegraphics[width=\linewidth]{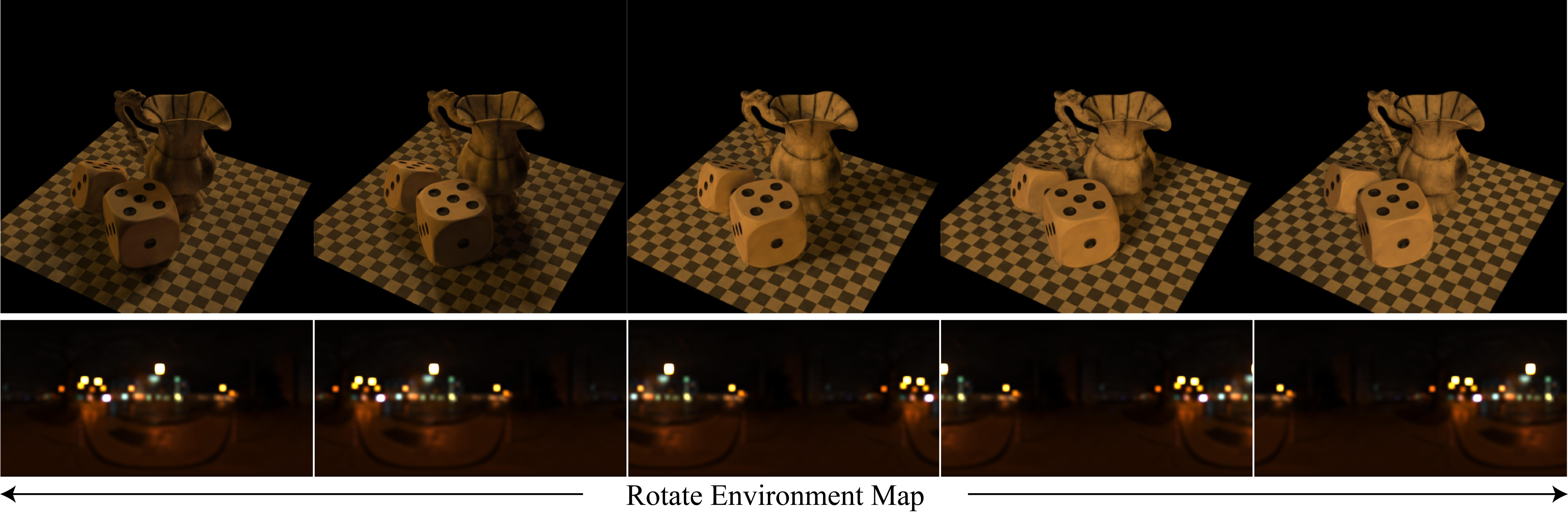}
    \caption{Illustration of relighting a scene using a sequence of rotated environment maps. The rendered results with global illumination are demonstrated on the first row, while the rotated environment maps are displayed on the second row.}
    \label{fig:env-map-relight}
\end{figure*}

\begin{figure}[thb]
    \centering
    \includegraphics[width=\linewidth]{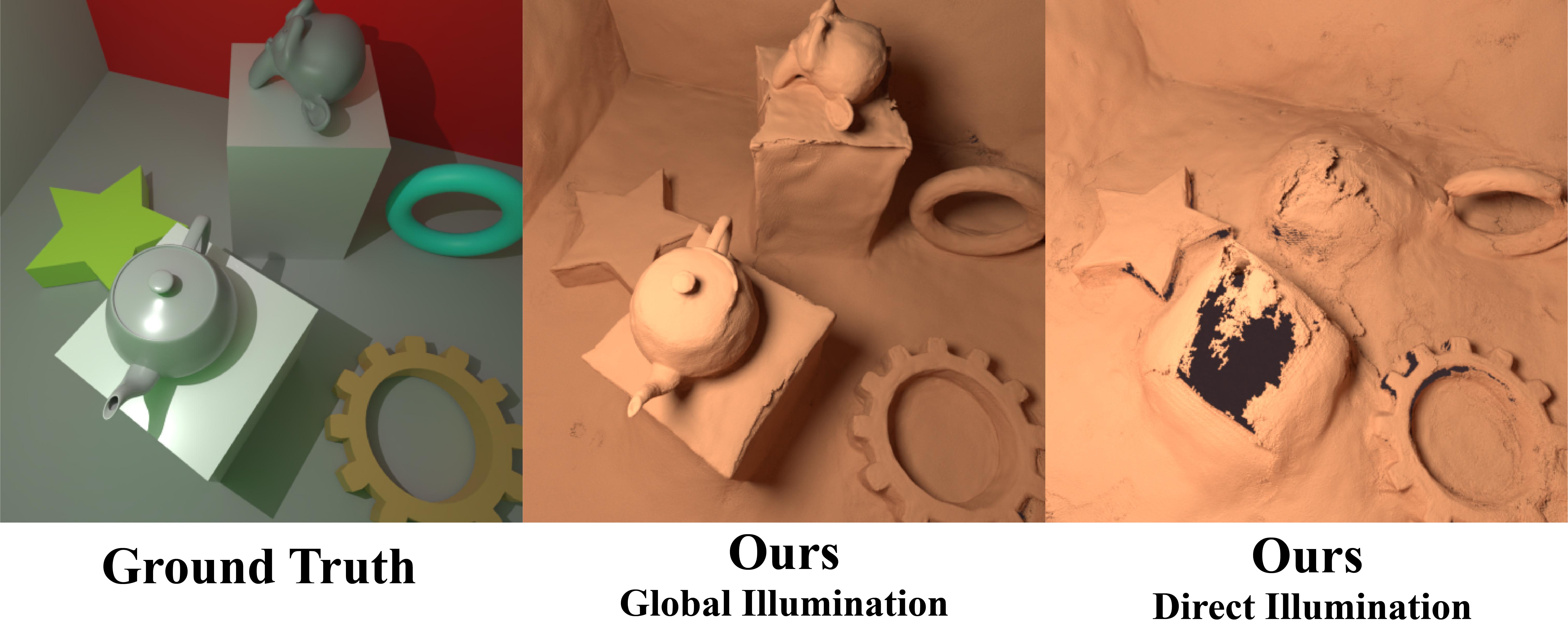}
    \caption{{Comparison between the optimizations using our method with direct illumination only and global illumination in terms of the geometry reconstruction quality on an enclosed box with objects inside.}}
    \label{fig:scene-toy}
\end{figure}

\paragraph{Scene decomposition.} Our method has the full understanding of light transport and is therefore capable of separating out different components of rendering. We use the ``CupPlane'' scene with non-metallic {and} rough materials 
provided by \citet{nrhints}. In Fig.~\ref{fig:material-edit}, we demonstrate the diffuse albedo and shininess components separately by passing the diffuse albedo and shininess instead of the calculated outgoing radiance into the rasterizer. 
We also separate out the direct illumination effects by running the progressive refinement only for $1$ step and indirect illumination effects, which is calculated as the difference between the full rendering result and the direct illumination effects.

\paragraph{Environment map relighting.} We further demonstrate the effectiveness of our method in relighting a scene using environment maps. The experiment is still conducted on the ``CupPlane'' scene. As shown in Fig.~\ref{fig:env-map-relight}, our approach produces high-quality relit results. As the environment map rotates, the resulting shadows adapt accordingly.

\paragraph{Preliminary experiment on an enclosed box.} {We provide a preliminary result on a synthetic enclosed box, featuring rich inter-reflection. Specifically, we make an enclosing box with $7$ objects, including one star, two cubes, one teapot, donuts, one gear and a monkey. We use $100$ training images, each of which is associated with a different point light source and the cameras are also put inside the box. As shown in Fig.~\ref{fig:scene-toy}, ours with direct illumination totally fails because the rich inter-reflection could not be handled, while ours with global illumination produces reasonable results.}
\section{Conclusions, Limitations and Future Work}
We propose the first work that derives differentiable global illumination on Gaussian surfels, which is an important theoretical development in its own right. Our work enables applications ranging from sparse-view relighting, material editing to accurate geometry reconstruction. We build the framework on top of the classic radiosity theory and therefore enable \emph{view-independent} rendering with both diffuse and specular effects. Notably, we extend the classic theory into non-binary visibility and semi-opaque primitives, propose novel solvers to efficiently solve the global illumination, deduce the efficient gradient calculation, and explore approximations to improve both efficiency and quality. Our method is also theoretically compatible with advancements in improving the differentiable rasterizer used in Gaussian splatting methods (e.g., \cite{hanson2024speedy, stopthepop}).

\begin{figure}[t]
    \centering
    \includegraphics[width=0.8\linewidth]{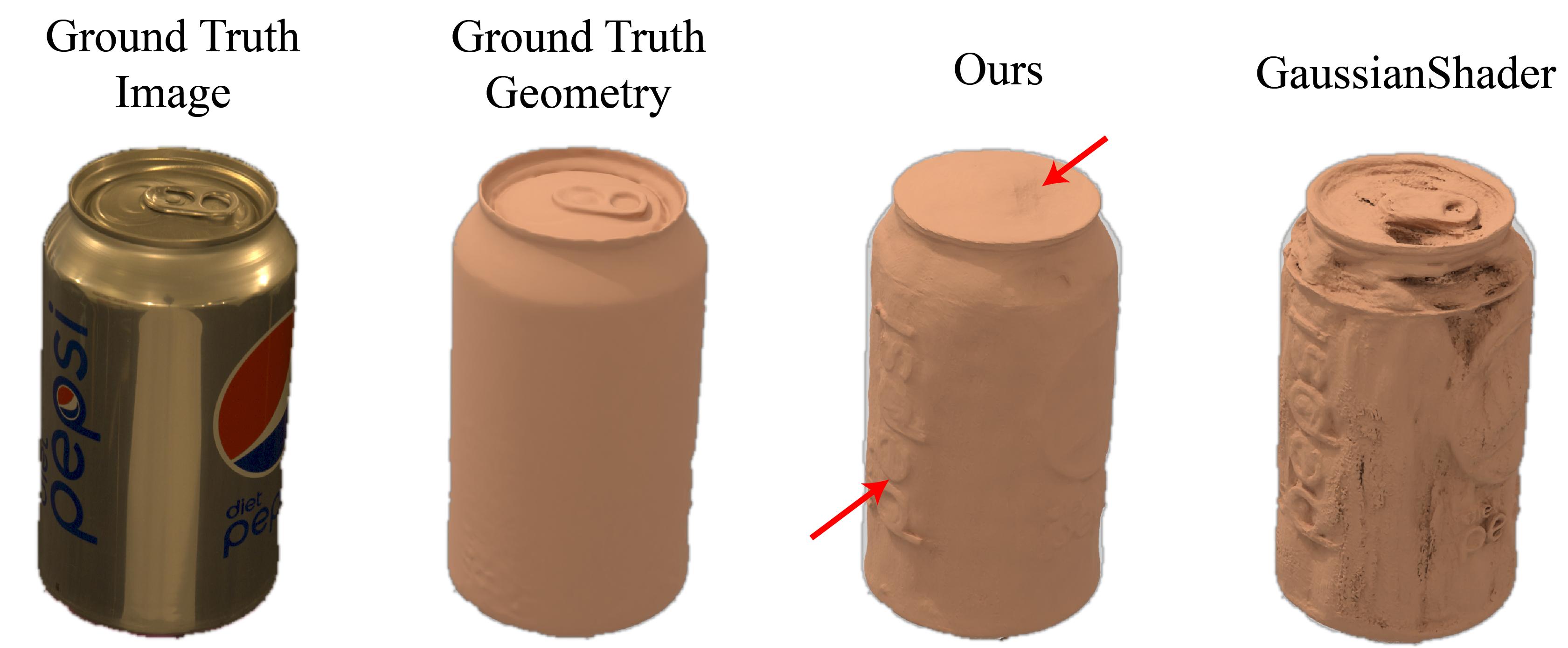}
    \caption{Illustration of a less successful case when the object is too shiny. Despite still being better than our baseline, ours exhibits obvious artifacts emphasized by the red arrows due to our reliance on finite degree of spherical harmonics basis to represent the material.}
    \label{fig:limitation-brdf}
\end{figure}

Our method is not without limitations. Even though the optimization should actively search for the best layout of Gaussian surfels, under the sparse view setting, the layout may not be optimal to catch all the details, evidenced by the shadows of the ``FurBall'' case in Fig.~\ref{fig:view-synthesis}{ and jaggy shadows in the accompanying video}. Besides, %
extending our framework into scene-level cases will be important as they feature more indirect illumination effects but this extension will also raise efficiency challenges{ and variance issues due to complicated occlusion, evidenced by the imperfect geometry reconstruction in Fig.~\ref{fig:scene-toy}}. {Speeding up our framework while maintaining low variance is of great importance in the future.} %
It will also be interesting to incorporate more complex material models, such as BSDFs or neural materials, to {lift the one-sided visibility assumption and} deal with objects with complex appearance such as the one shown in Fig.~\ref{fig:limitation-brdf} and the ``Translucent'' case in Fig.~\ref{fig:view-synthesis}.

We believe that our method paves the way for future explorations on efficient and differentiable physically-based rendering using discrete primitives, such as the Gaussian surfels.

\begin{acks}
We thank Rama Chellappa, Cheng Peng, Venkataram Sivaram, Haolin Lu, and Ishit Mehta for discussions. We thank the anonymous reviewers for detailed comments and suggestions. 
We also thank NVIDIA for GPU gifts.
This work was supported in part by the Intelligence Advanced Research Projects Activity (IARPA) via Department of Interior/ Interior Business Center (DOI/IBC) contract number 140D0423C0076. 
The views and conclusions contained herein are those of the authors and should not be interpreted as necessarily representing %
IARPA, DOI/IBC, or the U.S. Government. 
We also acknowledge support from ONR grant
N00014-23-1-2526, NSF grant 2127544 and 2238839, NSF grants 2100237 and 2120019 for the Nautilus cluster, gifts from Adobe, Google, Qualcomm and Rembrand, an NSERC Postdoctoral Fellowship, the Ronald L. Graham Chair and the UC San Diego Center for Visual Computing.
\end{acks}

\bibliographystyle{ACM-Reference-Format}
\bibliography{bibliography}

\setcounter{section}{0}
\renewcommand{\thesection}{\Alph{section}}
\section{BRDF details}
\label{app:brdf}
Given the BRDF defined in Eqn.~\eqref{eqn:brdf}, we have the closed-form solutions~\cite{sh} for the coefficient vector $\mathbf{f}^c$ as:
\begin{equation}
\begin{aligned}
    c_{00}&=4\mathbf{k} \mathbf{a}^d + (1-\mathbf{k}) \mathbf{a}^s \\
    c_{lm}&=(-1)^m (1-\mathbf{k}) \frac{(\mathbf{s}+1)(\mathbf{s}-1)\cdots(\mathbf{s}-l+2)}{(\mathbf{s}+2)(\mathbf{s}+4)\cdots(\mathbf{s}+l+1)}\mathbf{a}^s, \text{when } l\text{ is odd} \\
    c_{lm} &= (-1)^m (1-\mathbf{k}) \frac{\mathbf{s}(\mathbf{s}-2)\cdots(\mathbf{s}-l+2)}{(\mathbf{s}+3)(\mathbf{s}+5)\cdots(\mathbf{s}+l+1)}\mathbf{a}^s, \text{when }l\text{ is even} \\
\end{aligned}
\end{equation}
where the specular part of the BRDF is clamped to be $0$ at the invisible hemisphere, and we use up to $9$-degree spherical harmonics basis for reasonably glossy materials. 

\section{Forward rendering details}
\label{app:forward}
We start our derivation from the rendering equation \cite{rendering_equation} without using the {constant assumption} %
for the outgoing radiance, while the emission and BRDF are still constant within each kernel. We then incorporate the assumption regarding constant outgoing radiance within a kernel, and derive the solution to the underlying rendering equation using the weighted residual form. 
We start the discussion by assuming that there exists an outgoing direction $\bm{\omega}_O$ and then eventually generalize the discussion to the coefficient space of the spherical harmonics basis discussed above.

Specifically, from the rendering equation, we know that for the $i^\text{th}$ kernel:
\begin{equation}
\begin{aligned}
\label{eqn:rendering-equation}
    &\forall \mathbf{x}\in P_i, \\ 
    &L_O(\mathbf{x}, \bm{\omega}_O) =  {E}_i(\bm{\omega}_O)+\alpha_i(\mathbf{x})\int_{\bm{\Omega}} f_i(\bm{\omega}_I, \bm{\omega}_O)L_I(\mathbf{x}, \bm{\omega}_I)|\mathbf{n}_i\cdot\bm{\omega}_I|d\bm{\omega}_I, 
\end{aligned}
\end{equation}
where $\bm{\Omega}=\{\bm{\omega}_I:\bm{\omega}_I\cdot\mathbf{n}_i<0\}$ denotes the visible hemisphere, and $\alpha_i$ is multiplied because our primitive can be semi-transparent, which does not fully absorb all the energy.

By combining Eqn.~\eqref{eqn:L_I} and Eqn.~\eqref{eqn:rendering-equation}, we could expand $L_I$ as:
\begin{equation}
\label{eqn:before-swap}
\begin{aligned}
    \forall \mathbf{x}\in P_i, L_O(\mathbf{x}, \bm{\omega}_O) =&  {E}_i(\bm{\omega}_O)+\alpha_i(\mathbf{x})\int_{\bm{\omega}} f_i(\bm{\omega}_I, \bm{\omega}_O) |\mathbf{n}_i\cdot\bm{\omega}_I| \\&\underbrace{[\sum_{j=1}^{N}L_O(\mathbf{t}_j,\bm{\omega}_I)\alpha_j(\mathbf{t}_j)
    \prod_{k=1}^{j-1}(1-\alpha_k(\mathbf{t}_k))]}_{L_I(\mathbf{x},\bm{\omega}_I)}d\bm{\omega}_I
\end{aligned}
\end{equation}

\begin{figure}
    \centering
    \includegraphics[width=\linewidth]{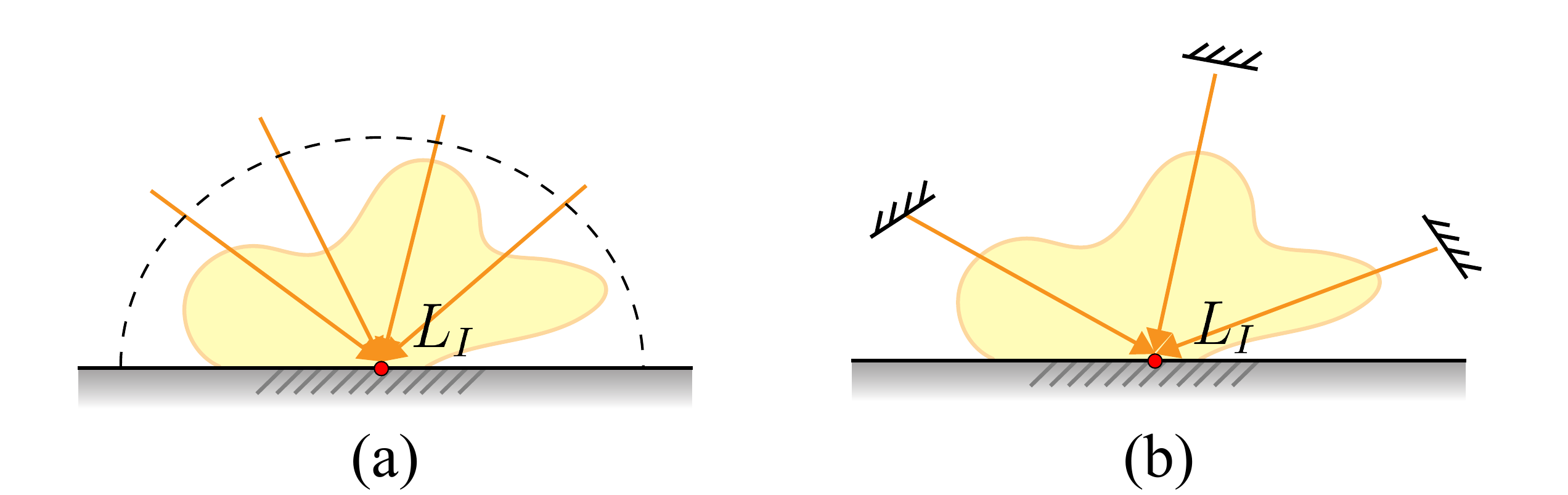}
    \caption{Illustration of swapping the integral and the summation in Eqn.~\ref{eqn:before-swap}. \textbf{(a)} The incident radiance at the red point is the integral of radiances coming from all potential directions. \textbf{(b)} After swapping the integral and the summation, the incident radiance at the red point is the summation of radiances coming from all other kernels in space.}
    \label{fig:incident-light}
\end{figure}

We then swap the integral and the summation. The key to interpret it is that, now we are choosing a specific direction first and then calculating how much the contribution is from all the kernels given that direction. It is equivalent to, say, choosing a specific kernel first and then calculating how much contribution it gives. A visualization is provided in Fig.~\ref{fig:incident-light}.

Therefore, we have:
\begin{equation}
\begin{aligned}
    \forall \mathbf{x}\in P_i, L_O(\mathbf{x}, \bm{\omega}_O) =&  {E}_i(\bm{\omega}_O)+\alpha_i(\mathbf{x})\sum_{j=1}^{N}\int_{P_j} \underbrace{f_i(\bm{\omega}_I, \bm{\omega}_O)}_\text{BRDF Response}\underbrace{L_O(\mathbf{t}_j,\bm{\omega}_I)}_\text{Radiance} \\
    &\underbrace{\alpha_j(\mathbf{t}_j)\prod_{k=1}^{j-1}(1-\alpha_k(\mathbf{t}_k))\frac{|\mathbf{n}_i\cdot\bm{\omega}_I||\mathbf{n}_j\cdot\bm{\omega}_I|}{||\mathbf{x}-\mathbf{x'}||_2^2}}_\text{Decay}d\mathbf{x'}, 
\end{aligned}
\end{equation}
where $\bm{\omega}_I=(\mathbf{x}-\mathbf{x'})/||\mathbf{x}-\mathbf{x'}||_2^2$. To simplify the notation, we define $v_{ji}(\mathbf{x}',\mathbf{x})$ as the decay component, which is part of the final fused decay component $V_{ji}$ we will introduce later. Therefore, we have:
\begin{equation}
\label{eqn:rendering-equation-kernel-summation}
\begin{aligned}
    L_O(\mathbf{x}, \bm{\omega}_O) = &{E}_i(\bm{\omega}_O) + \\ &\alpha_i(\mathbf{x})\sum_{j=1}^{N}\int_{P_j} {f_i(\bm{\omega}_I, \bm{\omega}_O)}{L_O(\mathbf{t}_j,\bm{\omega}_I)} {v_{ji}(\mathbf{x}', \mathbf{x})}d\mathbf{x'}, 
\end{aligned}
\end{equation}
for all $\mathbf{x}\in P_i$.

We then incorporate our {assumption that the material, emission and outgoing radiance are constant within the support}%
. Specifically, for the $i^\text{th}$ kernel, we intend to approximate the \emph{spatially variant} \emph{directionally-varying} outgoing radiance $L_O$ on the $i^\text{th}$ kernel with a \emph{spatially invariant} \emph{directionally-varying} quantity, denoted as $B_i$, which is still conditioned on the viewing direction. Accordingly, the \emph{spatially variant} $\alpha_i(\mathbf{x})$ is also approximated by the \emph{spatially invariant} $\alpha_i(\mathbf{p}_i)$.
The error of the approximation can be nicely captured by the \emph{weighted residual form} as in the radiosity theory. Specifically, we first replace the $L_O$ on both sides of Eqn.~\ref{eqn:rendering-equation-kernel-summation} with the defined spatially invariant directionally-varying quantity, and define the residual for the $i^\text{th}$ kernel as:
\begin{equation}
\forall\mathbf{x}\in P_i, r_i(\mathbf{x},\bm{\omega}_O) = {B}_i(\bm{\omega}_O)-E_i(\bm{\omega}_O)-\alpha_i\sum_{j=1}^{N}\int_{P_j}f_iB_jv_{ji}d\mathbf{x'}, 
\end{equation}
where input variables of the function are dropped for clarity.

The weighted residual form is then applied to aggregate the residual for all points on the tangent plane such that:
\begin{equation}
\label{eqn:weighted-residual-form}
    \sum_{i=1}^{N} \sum_{j=1}^{N}\int_{\mathbb{R}^3} \kappa_j(\mathbf{x})r_i(\mathbf{x}, \bm{\omega}_O)d\mathbf{x}=0,  
\end{equation}
where $\kappa_j$ is a weighting function that is non-zero only at $P_j$. $\kappa_j$ could be $\alpha_j$ again, which is then known as the \emph{Galerkin form}, or other weighting function, such as Gaussian, Dirac delta, \emph{etc.}

We then introduce our second and last assumption to easily solve Eqn.~\eqref{eqn:weighted-residual-form}: \textit{Any two kernels are non-overlapping.}

This assumption is also adopted in the splatting algorithm \cite{3DGS, 2DGS, volume-splatting} and ensured by the subdivision in the classic radiosity theory. However, this assumption could cause energy conservation issues, but we prefer the simplicity and, in our inverse rendering applications, the optimization helps prevent the energy explosion.

With this assumption, since $\kappa_j$ is only non-zero at $P_j$ and $r_i$ is non-zero at $P_i$, Eqn.~\ref{eqn:weighted-residual-form} can be simplified as:
\begin{equation}
    \int_{P_i}\kappa_i(\mathbf{x})r_i(\mathbf{x},\bm{\omega}_O)d\mathbf{x}=0, 
\end{equation}
which leads to:
\begin{equation}
\label{final-radiosity-equation}
    B_i(\bm{\omega}_O)=E_i(\bm{\omega}_O)+\frac{\alpha_i(\mathbf{p}_i)}{\Lambda_i}\sum_{j=1}^{N}\int_{P_i}\kappa_i(\mathbf{x})\int_{P_j}f_i{B}_jv_{ji}d\mathbf{x'}d\mathbf{x}, 
\end{equation}
where $\Lambda_i=\int_{P_i}\kappa_i(\mathbf{x}) d\mathbf{x}$. 
Familiar readers should identify the resemblance between this equation and the classical radiosity equation \cite{radiosity_sh} except that every element now is not fully opaque. Therefore, only part of the emitted radiance is absorbed. Furthermore, the visibility is non-binary, which facilitates the later discussed optimization. 

We then convert this equation into the coefficient space of the spherical harmonics basis. Specifically, we have:
\begin{equation}
\begin{aligned}
    \mathbf{B}^c_i &= \mathbf{E}^c_i + \frac{\alpha_i(\mathbf{p}_i)}{\Lambda_i}\sum_{j=1}^{N}\int_{P_i}\int_{P_j}{[\mathbf{f}^c_i\times \overline{\mathbf{Y}}(\bm{\omega}_I)]} {[\mathbf{Y}(\bm{\omega}_I)^T\mathbf{B}^c_j]} {V}_{ji} d\mathbf{x}' d\mathbf{x} \\
     &= \mathbf{E}^c_i + \frac{\alpha_i(\mathbf{p}_i)}{\Lambda_i}\sum_{j=1}^{N}\int_{P_i}\int_{P_j}{\mathbf{f}_i(\bm{\omega}_I)} {[\mathbf{Y}(\bm{\omega}_I)^T\mathbf{B}^c_j]} {V}_{ji} d\mathbf{x}' d\mathbf{x}, 
\end{aligned}
\end{equation}
where $V_{ji}=\kappa_i(\mathbf{x})v_{ji}$. This equation is exactly Eqn.~\eqref{eqn:before-final-radiosity}.

\section{Backpropagation details}
\label{app:backward}
We start the analysis by expanding Eqn.~\eqref{eqn:final-radiosity} into enumerating paths connecting kernels:
\begin{equation}
\label{eqn:final-radiosity-expanded}
\begin{aligned}
    \mathbf{B}^c_i =& \mathbf{E}^c_i + \sum_{j=1}^{N}\sum_{
    \begin{array}{c}
         \text{all paths }x_1,x_2,...,x_l \\ \text{ connecting }i\text{ and }j
    \end{array}}\int_{P_i}\int_{P_{x_1}}\int_{P_{x_1}}\cdots\int_{P_{x_l}}\int_{P_{x_l}}\int_{P_{j}} \\
    & \mathbf{f}_i{V}_{x_1i}(\mathbf{Y}^T\mathbf{f}_{x_1}) {V}_{x_2x_1}(\mathbf{Y}^T\mathbf{f}_{x_2})\cdots {V}_{x_lx_{l-1}}(\mathbf{Y}^T\mathbf{f}_{x_l}) {V}_{jx_l}(\mathbf{Y}^T\mathbf{E}^c_j)d\cdots d\mathbf{x}, 
\end{aligned}
\end{equation}
where $l$ denotes the length of the path. Notice that each kernel in the middle, i.e., $x_1,x_2,...x_l$, can also be emissive, and this form also accounts for their emissions. This form resembles path tracing except that, we explicitly discretize the whole space as a finite number of kernels and calculate the light transport among them.

\begin{figure}[t]
    \centering
    \includegraphics[width=\linewidth]{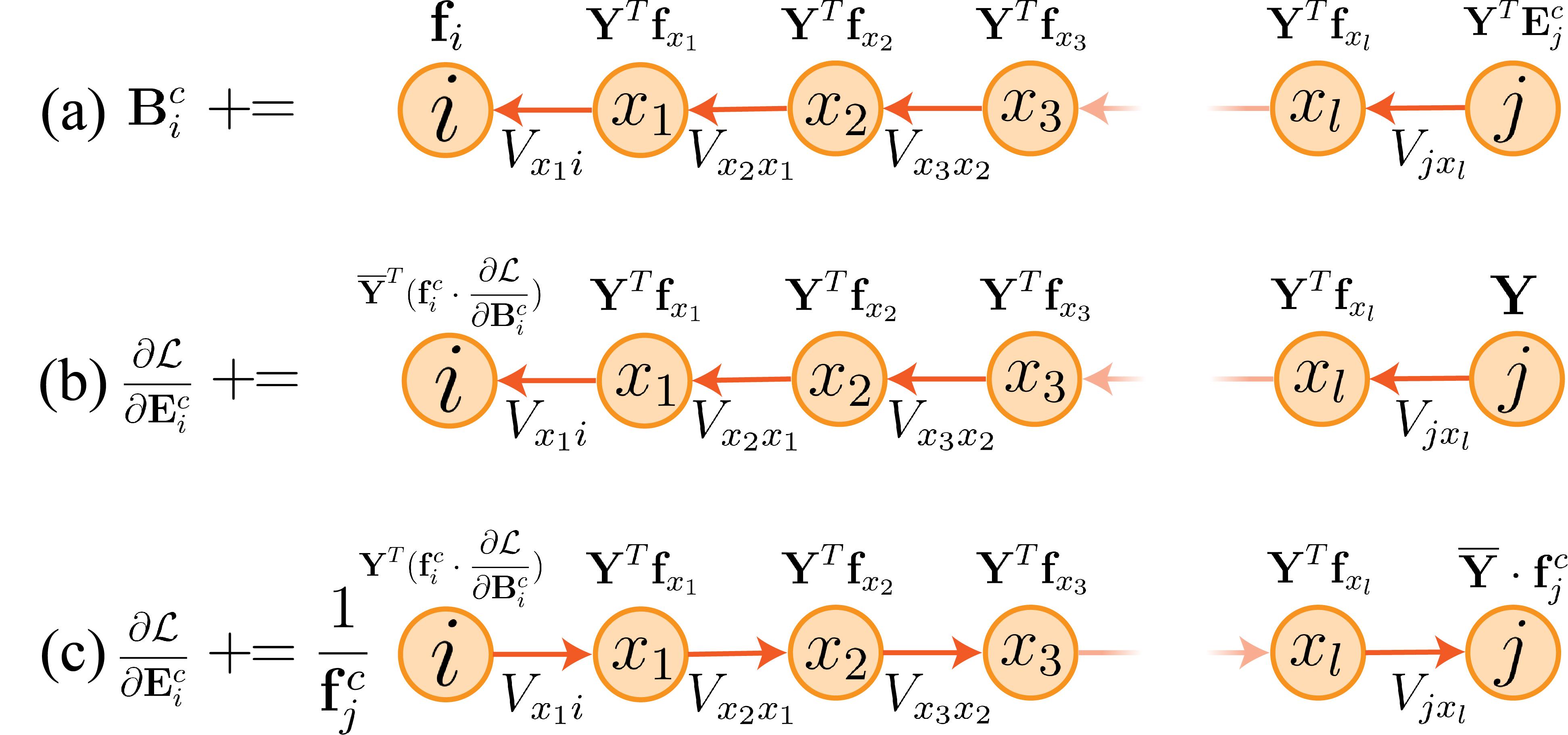}
    \caption{Illustration of calculating the partial derivatives with respect to the emissions by considering a single path $(x_1,x_2,...,x_l)$ connecting the $i^\text{th}$ and $j^\text{th}$ kernels. The arrow denotes the omitted direction for evaluating the spherical harmonics $\mathbf{Y}$ and BRDF $\mathbf{f}$. 
    \textbf{(a)} In the forward pass, the radiance $\mathbf{E}^c_j$ is emitted from the $j^\text{th}$ kernel along the direction connecting the $j^\text{th}$ and $x_l^\text{th}$ kernels, and is further processed through consecutive decay ($V$ term) and BRDF ($\mathbf{Y}^T\mathbf{f}$ term) to propagate into the $i^\text{th}$ kernel. 
    \textbf{(b)} In the backward pass, we need to reverse the forward procedure in (a) to propagate the weighted gradient $\mathbf{f}_i\times\frac{\partial\mathcal{L}}{\partial\mathbf{B}^c_i}$ from the $i^\text{th}$ kernel into the $j^\text{th}$ kernel. However, the direction cannot be directly reversed.
    \textbf{(c)} By noticing that $\mathbf{Y}(\bm{\omega})=\mathbf{\overline{Y}}(-\bm{\omega})$ and the reciprocal property of BRDF, i.e., $\mathbf{Y}^T(\bm{\omega}_O)\mathbf{f}_i(\bm{\omega}_I)=\mathbf{f}^T_i(-\bm{\omega}_O)\mathbf{Y}(-\bm{\omega}_I)$, we could then reverse the direction as well, which in turn makes the gradient propagation equivalent to a forward pass except that the weighted gradient is propagated instead of the emissions, and the asymmetric decay term $V$ is also flipped. }
    \label{fig:backward}
\end{figure}

\begin{figure*}
    \centering
    \includegraphics[width=\linewidth]{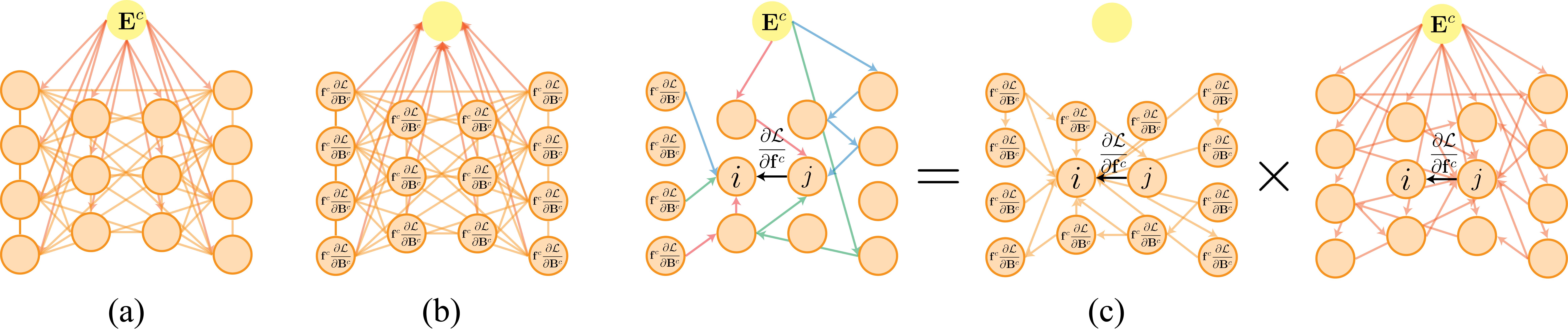}
    \caption{Conceptual illustration of the forward and backward pass for the differentiable light transport. \textbf{(a)} In the forward pass, emission $E$ from the light source is propagated in the scene to each kernel to calculate the radiosity $\mathbf{B}^c$. \textbf{(b)} In the backward pass, weighted gradient $\mathbf{f}^c\times\frac{\partial\mathcal{L}}{\partial\mathbf{B}^c}$ of each kernel is propagated in the scene to calculate the $\partial\mathcal{L}/\partial\mathbf{E}^c$. \textbf{(c)} To calculate the gradient with respect to the ``bridge'', or specifically the decay term and the BRDF term between any two kernels, we need to consider each path passing through such bridge. We take the $\partial\mathcal{L}/\partial\mathbf{f}^c$ on the bridge connecting the $i^\text{th}$ kernel and the $j^\text{th}$ kernel as an example here. For each path, the gradient can be roughly described as the product of the propagated weighted gradient $\mathbf{f}^c\times\frac{\partial\mathcal{L}}{\partial\mathbf{B}^c}$ and the emission $E$. Notice that, by considering every path, the propagated weighted gradient needs to be multiplied with every propagated emission, which is exactly equivalent to the forward pass in (a). Similarly, the propagated emission needs to be multiplied with every propagated weighted gradient, which is exactly equivalent to the backward pass in (b). Therefore, the $\partial\mathcal{L}/\partial\mathbf{f}^c$ can be effectively analytically defined using the $\mathbf{B}^c$ and $\partial\mathcal{L}/\partial\mathbf{E}^c$ calculated in (a) and (b).}
    \label{fig:forward-backward}
\end{figure*}

\paragraph{Gradients with respect to the emissions.} We first consider the partial derivatives for the emissions as they will be used in calculating other gradients as well. 
From Eqn.~\eqref{eqn:final-radiosity-expanded}, we first consider a single path $(x_1,x_2,...,x_l)$ connecting the $i^\text{th}$ and $j^\text{th}$ kernels as shown in Fig.~\ref{fig:backward} (a). We can then derive the partial derivatives for the emission for a single path by reversing the forward procedure as explained in Fig.~\ref{fig:backward} (b):
\begin{equation}
\label{eqn:dL_dE_single_pass}
\begin{aligned}
    \frac{\partial\mathcal{L}}{\partial\mathbf{E}^c_j} =& \int_{P_i}\int_{P_{x_1}}\int_{P_{x_1}}\cdots\int_{P_{x_l}}\int_{P_{x_l}}\int_{P_{j}}(\mathbf{f}_i^T \frac{\partial\mathcal{L}}{\partial\mathbf{B}^c_i}) \\
    & {V}_{x_1i}(\mathbf{Y}^T\mathbf{f}_{x_1}) {V}_{x_2x_1}(\mathbf{Y}^T\mathbf{f}_{x_2})\cdots {V}_{x_lx_{l-1}}(\mathbf{Y}^T\mathbf{f}_{x_l}){V}_{jx_l} \mathbf{Y}d\cdots d\mathbf{x}. 
\end{aligned}
\end{equation}

Notice that $\mathbf{f}_i=\mathbf{f}_i^c\times\overline{\mathbf{Y}}$, and $\mathbf{Y}^T(\bm{\omega}_O)\mathbf{f}_i(\bm{\omega}_I)=\mathbf{f}^T_i(-\bm{\omega}_O)\mathbf{Y}(-\bm{\omega}_I)$ due to the \emph{reciprocal} property of the BRDF. We could rewrite Eqn.~\ref{eqn:dL_dE_single_pass} as explained in Fig.~\ref{fig:backward} (c):
\begin{equation}
\begin{aligned}
    \frac{\partial\mathcal{L}}{\partial\mathbf{E}^c_j} =& \frac{1}{\mathbf{f}^c_j}\int_{P_j}\int_{P_{x_l}}\int_{P_{x_l}}\cdots\int_{P_{x_1}}\int_{P_{x_1}}\int_{P_{i}} \\
    & {\mathbf{f}_j} \overline{{V}_{x_lj}}(\mathbf{Y}^T{\mathbf{f}_{x_l}}) \cdots \overline{{V}_{x_{1}x_{2}}}(\mathbf{Y}^T{\mathbf{f}_{x_1}}) \overline{{V}_{ix_1}} (\mathbf{Y}^T({\mathbf{f}^c_i}\times \frac{\partial\mathcal{L}}{\partial\mathbf{B}^c_i}))d\cdots d\mathbf{x}. 
\end{aligned}
\end{equation}

Therefore, by accumulating every such single path and swapping $i$ and $j$ for symbol consistency, we have:
\begin{equation}
\label{eqn:dL_dE_final}
\begin{aligned}
    \frac{\partial\mathcal{L}}{\partial\mathbf{E}^c_i} =& \frac{\partial\mathcal{L}}{\partial\mathbf{B}^c_i} + \frac{1}{\mathbf{f}^c_i} \sum_{j=1}^{N}\sum_{
    \begin{array}{c}
         x_1,x_2,...,x_l
    \end{array}
    }\int_{P_i}\int_{P_{x_1}}\int_{P_{x_1}}\cdots\int_{P_{x_l}}\int_{P_{x_l}}\int_{P_{j}} \\
    & {\mathbf{f}_i} \overline{{V}_{x_1i}}(\mathbf{Y}^T{\mathbf{f}_{x_1}}) \cdots \overline{{V}_{x_{l}x_{l-1}}}(\mathbf{Y}^T{\mathbf{f}_{x_l}}) \overline{{V}_{jx_l}} (\mathbf{Y}^T({\mathbf{f}^c_j}\times \frac{\partial\mathcal{L}}{\partial\mathbf{B}^c_j}))d\cdots d\mathbf{x},  
\end{aligned}
\end{equation}
where $\overline{V_{ij}}=V_{ji}$. Note that this equation resembles the rewritten form (Eqn.~\eqref{eqn:final-radiosity-expanded}) of Eqn.~\eqref{eqn:final-radiosity}, except that the decay term is reversed and we are propagating the weighted gradient $\mathbf{f}^c_j\times\frac{\partial\mathcal{L}}{\partial\mathbf{B}^c_j}$ instead of emission.
By converting Eqn.~\eqref{eqn:dL_dE_final} into the form of Eqn.~\eqref{eqn:final-radiosity}, we get the exact Eqn.~\eqref{eqn:final_dL_dE}.
This relationship implies the duality between the forward rendering and gradient calculation. %
We provide a conceptual illustration of forward and backward pass in Fig.~\ref{fig:forward-backward} (a) and (b).

\paragraph{Gradients with respect to the BRDF}
From Eqn.~\eqref{eqn:final-radiosity-expanded}, we could also derive the partial derivative with respect to the $\mathbf{f}^c_i$. ${\partial\mathcal{L}}/{\partial\mathbf{f}^c_i}$ should accumulate the gradients from every path which passes through the $i^\text{th}$ kernel. We now present a high-level derivation of such gradients. As shown in Fig.~\ref{fig:forward-backward} (c), the contribution to the gradients from each path can be roughly described as the product of the propagated weighted gradient from the end kernel and the propagated emissions from the start kernel. Or, for each propagated weighted gradient, it should multiply with the accumulated propagated emissions from \emph{every} kernel, or $\mathbf{B}^c_j,\forall j\in\{1,2,...,N\}$ exactly. And, the further accumulation of each propagated weighted gradient leads to the exact definition of ${\partial\mathcal{L}}/{\partial \mathbf{E}^c_i}$. Namely, we have:
\begin{equation}
\label{eqn:dl-dbrdf}
\begin{aligned}
    \frac{\partial\mathcal{L}}{\partial\mathbf{f}^c_i} &= \frac{\partial\mathcal{L}}{\partial\mathbf{E}^c_i}\times\sum_{j=1}^N \int_{P_i}\int_{P_j} 
\overline{\mathbf{Y}}[{V}_{ji} (\mathbf{Y}^T \mathbf{B}^c_j) ] d\mathbf{x}' d\mathbf{x} \\
&=\frac{\partial\mathcal{L}}{\partial\mathbf{E}^c_i}\times \frac{(\mathbf{B}^c_i-\mathbf{E}^c_i)}{\mathbf{f}^c_i}, 
\end{aligned}
\end{equation}
which is an analytical non-recursive closed form solution since $\mathbf{B}^c_j$ sums all the propagated radiances passing through the $j^\text{th}$ kernel and ${\partial\mathcal{L}}/{\partial\mathbf{E}^c_i}$ sums all the propagated weighted gradient passing through the $i^\text{th}$ kernel. This equation is then the exact Eqn.~\eqref{eqn:final-dl-dbrdf}.

\paragraph{Gradients with respect to the geometry}
The remaining gradients are the partial derivatives with respect to the geometric properties (i.e., central points, scaling, \emph{etc.}) from the decay term and directional effects. The derivation for their analytical non-recursive forms is similar to the derivation of $\frac{\partial\mathcal{L}}{\partial\mathbf{f}^c_i}$.
As in the main paper, we define $\beta_i$ as a placeholder for the geometric property of the $i^\text{th}$ kernel that we want to differentiate through.

Similar to the derivation of partial derivatives with respect to the $\mathbf{f}^c_i$, we first consider every path that passes through a decay term $V$ that contains the $i^\text{th}$ kernel, and the gradient with respect to the decay term is the product of the propagated emissions from the start kernel and the propagated weighted gradient from the end kernel. For each propagated weighted gradient, it should multiply with the accumulated propagated emissions from \emph{every} kernel, or $\mathbf{B}^c_j,\forall j\in\{1,2,...,N\}$ exactly. And, the further accumulation of each propagated weighted gradient leads to the exact definition of $\partial\mathcal{L}/\partial\mathbf{E}^c_i$. Therefore, to calculate the gradients from every path passing through the decay term that contains the $i^\text{th}$ kernel, we need to enumerate every pair of kernels forming a decay term. We then reach Eqn.~\eqref{eqn:dl-dgeo-0}, where the $s^u_as^v_as^u_bs^v_b$ term comes from differentiating through the integration domain or boundary.

Besides, we also need to take the directional effects into account. The gradients with respect to the direction are similar to the gradients with respect to the decay term, but we only need to enumerate every possible pair of kernels, one of which is the $i^\text{th}$ kernel. We then reach Eqn.~\eqref{eqn:dl-dgeo-1}, where the gradients are composed of two symmetric terms, which come from evaluating the outgoing radiance and BRDF response with respect to a specific direction.

\section{Monte-Carlo Solver Details}
\label{app:mc}
We further generalize the proposed Monte-Carlo solver in Sec.~\ref{method-solver} to obtain a more efficient solver. Specifically, there is no constraint that the time step needs to be synchronized for all kernels. As we gradually update the estimated outgoing radiance for every kernel, we track its variance over time, and perform more steps for kernels with high variance. In practice, at each step, we select $N$ elements from $\{1,2,...,N\}$ with replacement and with probability proportional to the tracked variance instead of simply using $\{1,2,...,N\}$ (Line~\ref{alg:mc-select} in Alg.~\ref{monte-carlo-solver}).

Besides, we also improve the efficiency of next event estimation by grouping kernels into different groups. Formally, we rewrite the summation in Eqn.~\eqref{eqn:final-radiosity} as a double summation:
\begin{equation}
\begin{aligned}
    \mathbf{B}^c_i &= \mathbf{E}^c_i + \sum_\text{groups}\sum_{j\text{ in group}}\int_{P_i}\int_{P_j}{\mathbf{f}_i(\bm{\omega}_I)} {[\mathbf{Y}(\bm{\omega}_I)^T\mathbf{B}^c_j]} {V}_{ji} d\mathbf{x}' d\mathbf{x}, 
\end{aligned}
\end{equation}
where we use K-Means to group kernels into a preset number of groups with the averaged central points and normal directions and the summed current estimation of outgoing radiance as the proxy for importance sampling.

Therefore, we first sample a group and then sample a kernel within that group based on Eqn.~\eqref{eqn:importance-sampling} for the Monte-Carlo solver. This is effectively a very shallow tree with depth $2$. By re-writing the single summation into double summation, the complexity is reduced from $\mathcal{O}(N^2)$ into $\mathcal{O}(N\sqrt{N})$. Although the optimal is to rewrite the summation into $\ln(N)$ summations (i.e., tree with depth $\ln(N)$), we find that using double summation already has good performance in practice. We note that such an idea is conceptually similar to the established many-lights rendering~\cite{light-cut, stochastic-lightcut}. However, it is different from the established hierarchical methods (e.g., \cite{radiosity_hierarchical, radiosity_hierarchical_glossy}) in the radiosity theory where they group kernels such that the transmitted radiance is approximately identical for every element in the group. In our case, we still need to calculate the transmitted radiance for every kernel separately due to the non-binary visibility that makes the approximation harder.

\section{Approximation Details}
\label{app:appr}
In the approximation (Sec.~\ref{method-approx}), we need to evaluate the integral $\int_{P_j}\alpha_jd\mathbf{x}$. From \citet{gfsgs}, 
\begin{equation}
    \alpha_j(\mathbf{x})\approx 1-\exp(-0.03279(g_j\exp(-\frac{1}{2}(u^2+v^2)))^{3.4}), 
\end{equation}
where $u,v$ denote the coordinates of $\mathbf{x}$ in the local tangent plane $P_j$ of the $j^\text{th}$ kernel as defined in Eqn.~\ref{eqn:G}.

Therefore, we could transform the integral of $\alpha_j$ over the $P_j$ into the integral over the local tangent plane as:
\begin{equation}
\begin{aligned}
    &\int_{P_j} \alpha_j(\mathbf{x})d\mathbf{x} \\
    &= s^u_js^v_j \int_{0}^\infty\int_0^\infty 1-\exp(-0.03279(g_j\exp(-\frac{1}{2}(u^2+v^2)))^{3.4}) dudv \\
    &= 2\pi s^u_js^v_j \int_{0}^\infty r(1-\exp(-0.03279(g_j\exp(-\frac{1}{2}r^2))^{3.4})) dr \\
    &= 2\pi s^u_js^v_j \int_{0}^\infty 1-\exp(-0.03279g_j^{3.4}\exp(-3.4x)) dx \\
    &= \frac{2\pi}{3.4} s^u_js^v_j \int_{0}^\infty 1-\exp(-0.03279g_j^{3.4}\exp(-x)) dx \\
    &= \frac{2\pi}{3.4} s^u_js^v_j \int_{0}^\infty 1-\exp(-c_j\exp(-x)) dx \\
    &= \frac{2\pi}{3.4} s^u_js^v_j \int_{0}^1 \frac{1-\exp(-c_jy)}{y} dy \\
    &= \frac{2\pi}{3.4} s^u_js^v_j \int_{0}^{c_j} \frac{1-\exp(-y)}{y} dy
\end{aligned}
\end{equation}
where $c_j=0.03279g_j^{3.4}$ and $\int_0^{c_j}\frac{1-\exp(-y)}{y}dy$ is related to a specific function, i.e., the exponential integral. Therefore, we have the closed-form solution for $\int_{P_j}\alpha_jd\mathbf{x}$.

\end{document}